\documentclass{aa}

\usepackage[font=small,labelfont=bf]{caption}
\usepackage{newunicodechar}
\usepackage{xcolor}
\usepackage{orcidlink}
\usepackage{graphicx}
\usepackage{txfonts}
\usepackage{lipsum}
\usepackage{subcaption}         
\usepackage{lscape}             
\usepackage{placeins}           

\usepackage{booktabs}
\usepackage{multirow}
\usepackage[flushleft]{threeparttable}
                                
\usepackage{hyperref}

\DeclareRobustCommand{\okina}{%
  \raisebox{\dimexpr\fontcharht\font`A-\height}{%
    \scalebox{0.8}{`}%
  }%
}
\newunicodechar{ʻ}{\okina}

\begin{document}

   \title{EPRV Asteroseismology of the K dwarfs GJ~570~A and GJ~783~A\\ with ESPRESSO}

   \subtitle{Empirical mode-amplitude scaling in tension with 3D convection simulations}

   \titlerunning{EPRV Asteroseismology of the K dwarfs GJ~570~A and GJ~783~A}

   \authorrunning{T. L. Campante et al.}


   \author{
        T. L. Campante\inst{\ref{IA}}\fnmsep\inst{\ref{FCUP}}\orcidlink{0000-0002-4588-5389}
        \and
        Y. Zhou\inst{\ref{Oslo}}\orcidlink{0000-0003-0817-6126}
        \and
        Y. Li\inst{\ref{Hawaii}}\orcidlink{0000-0003-3020-4437}
        \and
        H. Kjeldsen\inst{\ref{Aarhus}}\orcidlink{0000-0002-9037-0018}
        \and
        R. R. Costa\inst{\ref{IA}}\fnmsep\inst{\ref{FCUP}}\orcidlink{0009-0008-6039-6381}
        \and
        A. W. Neitzel\inst{\ref{IA}}\fnmsep\inst{\ref{FCUP}}\orcidlink{0000-0001-6283-907X}
        \and
        J. H. C. Martins\inst{\ref{IA}}\orcidlink{0000-0002-1532-9082}
        \and\\
        A. M. Silva\inst{\ref{IA}}\fnmsep\inst{\ref{FCUP}}\orcidlink{0000-0003-4920-738X}
        \and
        V. Adibekyan\inst{\ref{IA}}\fnmsep\inst{\ref{FCUP}}\orcidlink{0000-0002-0601-6199}
        \and
        J. Gomes da Silva\inst{\ref{IA}}\orcidlink{0000-0001-8056-9202}
        \and
        J. R. Larsen\inst{\ref{Aarhus}}\orcidlink{0009-0006-0423-2353}
        \and
        S. G. Sousa\inst{\ref{IA}}\fnmsep\inst{\ref{FCUP}}\orcidlink{0000-0001-9047-2965}
        \and
        T. R. Bedding\inst{\ref{Sydney}}\orcidlink{0000-0001-5222-4661}
        \and\\
        D. Bossini\inst{\ref{Padova}}\orcidlink{0000-0002-9480-8400}
        \and
        W. J. Chaplin\inst{\ref{Bham}}\orcidlink{0000-0002-5714-8618}
        \and
        E. Corsaro\inst{\ref{Catania}}\orcidlink{0000-0001-8835-2075}
        \and
        M. S. Cunha\inst{\ref{IA-Coimbra}}\orcidlink{0000-0001-8237-7343}
        \and
        M. Deal\inst{\ref{LUPM}}\orcidlink{0000-0001-6774-3587}
        \and
        D. Huber\inst{\ref{Hawaii}}\orcidlink{0000-0001-8832-4488}
        \and
        M. N. Lund\inst{\ref{Aarhus}}\orcidlink{0000-0001-9214-5642}
        \and\\
        M. S. Lundkvist\inst{\ref{Aarhus}}\orcidlink{0000-0002-8661-2571}
        \and
        T. S. Metcalfe\inst{\ref{WDRC}}\orcidlink{0000-0003-4034-0416}
        \and
        M. J. P. F. G. Monteiro\inst{\ref{IA}}\fnmsep\inst{\ref{FCUP}}\orcidlink{0000-0003-0513-8116}
        \and
        N. C. Santos\inst{\ref{IA}}\fnmsep\inst{\ref{FCUP}}\orcidlink{0000-0003-4422-2919}
        }

   \institute{
        Instituto de Astrof\'{\i}sica e Ci\^{e}ncias do Espa\c{c}o, Universidade do Porto,  Rua das Estrelas, 4150-762 Porto, Portugal\\
        \email{tiago.campante@astro.up.pt}
        \label{IA}
   \and
        Departamento de F\'{\i}sica e Astronomia, Faculdade de Ci\^{e}ncias da Universidade do Porto, Rua do Campo Alegre, s/n, 4169-007 Porto, Portugal
        \label{FCUP}
    \and
        Rosseland Centre for Solar Physics (RoCS), Institute of Theoretical Astrophysics, University of Oslo, PO Box 1029, Blindern, 0315 Oslo, Norway
        \label{Oslo}
    \and
        Institute for Astronomy, University of Hawai\okina i, 2680 Woodlawn Drive, Honolulu, HI 96822, USA
        \label{Hawaii}
    \and
        Stellar Astrophysics Centre (SAC), Department of Physics and Astronomy, Aarhus University, Ny Munkegade 120, 8000 Aarhus C, Denmark
        \label{Aarhus}
    \and
        Sydney Institute for Astronomy (SIfA), School of Physics, University of Sydney, NSW 2006, Australia
        \label{Sydney}
    \and
        INAF -- Osservatorio Astronomico di Padova, Vicolo dell'Osservatorio 5, 35122 Padova, Italy
        \label{Padova}
    \and
        School of Physics and Astronomy, University of Birmingham, Edgbaston, Birmingham B15 2TT, UK
        \label{Bham}
    \and
        INAF -- Osservatorio Astrofisico di Catania, Via S.~Sofia 78, 95123 Catania, Italy
        \label{Catania}
    \and
        Instituto de Astrof\'{\i}sica e Ci\^{e}ncias do Espa\c{c}o, Departamento de F\'{\i}sica, Universidade de Coimbra, 3004-516 Coimbra, Portugal
        \label{IA-Coimbra}
    \and
        LUPM, Universit\'e de Montpellier, CNRS, Place Eug\`ene Bataillon, 34095 Montpellier, France
        \label{LUPM}
    \and
        Center for Solar-Stellar Connections, WDRC, 9020 Brumm Trail, Golden, CO 80403, USA
        \label{WDRC}
        }

   \date{Received Month Day, 2026}

 
  \abstract
   {Solar-like oscillations in K dwarfs have intrinsically low amplitudes, making them challenging to detect in space-based photometry. The advent of Extremely Precise Radial Velocity (EPRV) instrumentation enables Doppler asteroseismology in the mid-K regime, providing a new opportunity to test mode-excitation and damping physics.}
   {We investigate the scaling of oscillation amplitudes with stellar parameters and assess whether 3D surface convection simulations reproduce the observed amplitude behavior.}
   {High-cadence ESPRESSO radial-velocity time series were obtained over five and six consecutive nights for the K dwarfs GJ~570~A (K4~V) and GJ~783~A (K2.5~V), respectively. We identified oscillation modes from weighted power spectra and extracted them via sine-wave fitting following spectral deconvolution. Peak mode amplitudes were measured from heavily smoothed power spectra. For comparison, we performed 3D radiative-hydrodynamical simulations including excitation and damping-rate calculations.}
   {We measured 6 modes for GJ~570~A and 24 for GJ~783~A ($\ell\!=\!0$--$2$, and $\ell\!=\!3$ for GJ~783~A). The peak radial-mode amplitudes are $2.57 \pm 0.52\:{\rm cm\,s^{-1}}$ and $4.90 \pm 0.81\:{\rm cm\,s^{-1}}$, respectively. Combined with previous Doppler detections, these results support a transition from an $\sim\!(L/M)^{0.7}$ amplitude scaling to a steeper $\sim\!(L/M)^{1.5}$ scaling, occurring near the G/K-dwarf boundary toward decreasing $L/M$. In contrast, 3D simulations predict an $\sim\!(L/M)^{0.9}$ scaling for K dwarfs, revealing tension between observations and first-principles convection models.}
   {EPRV asteroseismology extends into the mid-K regime and exposes a steep mode-amplitude scaling that our 3D convection simulations do not reproduce. This discrepancy points to missing physics in our implementation of convection–pulsation coupling and/or limitations in the adopted modeling assumptions, motivating further theoretical investigation.}

   \keywords{Asteroseismology --
            Stars: individual: GJ~570~A, GJ~783~A --
            Stars: late-type --
            Stars: oscillations (including pulsations) --
            Techniques: radial velocities
        }

   \maketitle
   \nolinenumbers

\section{Introduction}\label{sect:intro}
\begin{figure}[!t]
\centering
\includegraphics[width=0.5\textwidth,trim=0.35cm 0.55cm 0cm 0cm,clip]{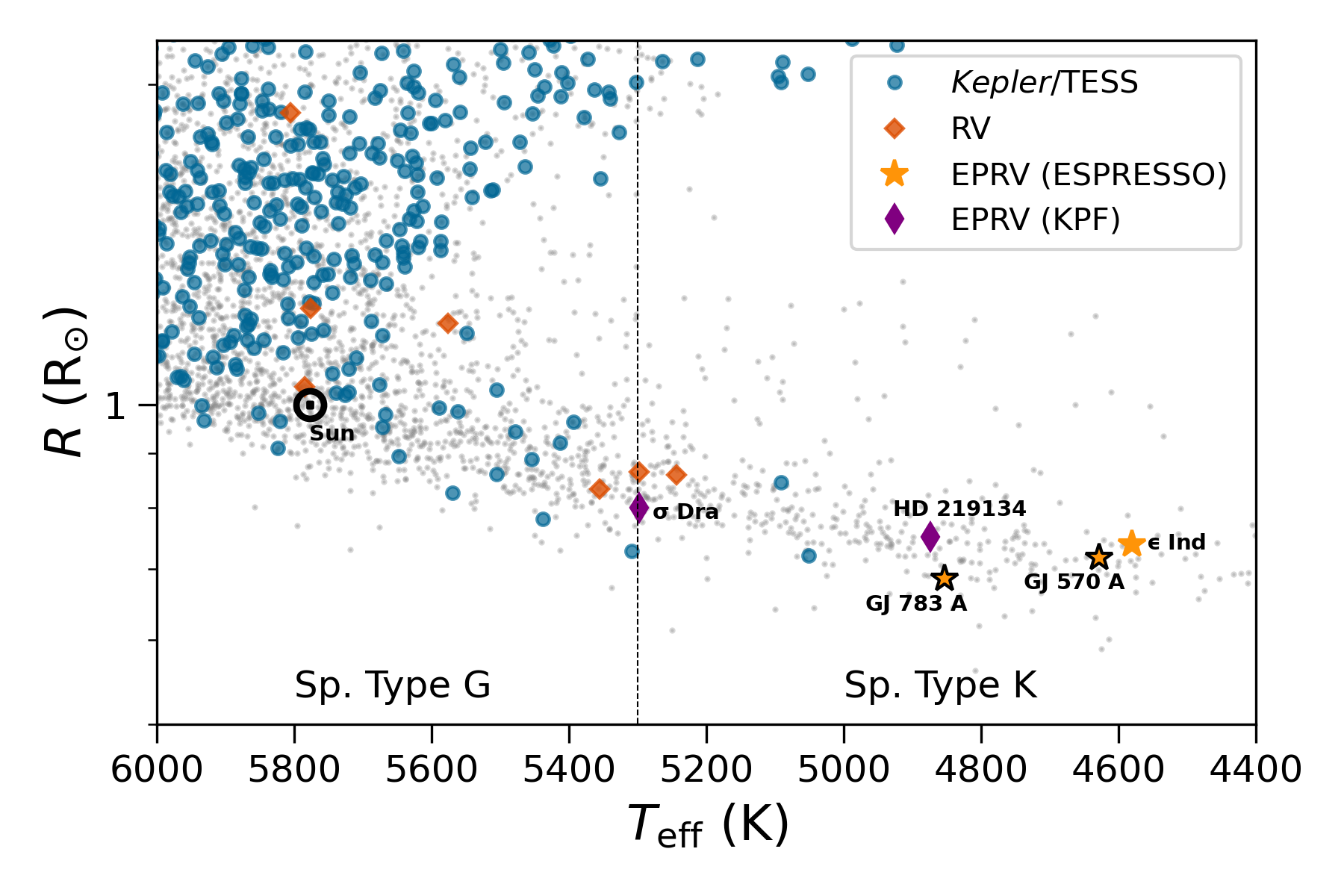}
\caption{Stellar radius--effective temperature diagram highlighting asteroseismic detections from {\it Kepler} and TESS photometry \citep[blue circles;][]{Mathur17,Hatt23,Lund25}, earlier RV campaigns \citep[red diamonds; see][and references therein]{Arentoft08,Kjeldsen08}, and recent EPRV observations of K dwarfs with ESPRESSO \citep[orange star symbols;][]{Campante24} and KPF \citep[purple thin diamonds;][]{Hon24,Li25}. The two stars analyzed in this work, GJ~570~A and GJ~783~A, are highlighted among the ESPRESSO sample. The stellar background sample (gray dots) is taken from the TESS Input Catalog \citep[TIC;][]{TIC}. The Sun is denoted by its usual symbol. Spectral types G and K are delimited by a vertical dashed line.}
\label{fig:HRD}
\end{figure}

Notable progress in asteroseismology has been made due to missions such as Convection, Rotation and planetary Transits \citep[CoRoT;][]{CoRoT}, {\it Kepler}/K2 \citep[][]{Kepler,K2}, and the Transiting Exoplanet Survey Satellite \citep[TESS;][]{TESS}. These missions have delivered high-precision photometric data from space, allowing for in-depth investigations of the internal structures of solar-type and red-giant stars, which exhibit solar-like oscillations driven by convection \citep[][]{ChaplinMiglio13,Aerts21}. Despite this, the small-amplitude solar-like oscillations in K dwarfs \citep[amplitudes below a few parts per million or, correspondingly, $10\:{\rm cm\,s^{-1}}$;][]{Kjeldsen08,Verner11} remain extremely challenging to detect in space-based photometry \citep{Mathur17,Hatt23}. This difficulty persists even with access to multi-year {\it Kepler} \citep[e.g.,][]{Kepler444,Lundkvist16} and multi-sector TESS \citep[e.g.,][]{Campante16,Hon24,Lund25} observations. As a result, only a handful of dwarfs cooler than the Sun show oscillations in photometry, with detections dropping off sharply below ${\sim} 5500\:{\rm K}$ (see Fig.~\ref{fig:HRD}).

Doppler velocity observations offer a viable alternative to space-based photometry. Stellar ``noise'' from activity and granulation is significantly lower in Doppler than in photometry \citep{Harvey88,SONG}, resulting in a higher signal-to-noise ratio (S/N) over the typical oscillation timescales \citep{Hon24,Kjeldsen25,Adam26}. This motivated several pre-{\it Kepler} radial-velocity (RV) campaigns targeting late-G/early-K dwarfs \citep[e.g.,][]{Kjeldsen05,CarrierEggenberger06,Teixeira09}, using then state-of-the-art spectrographs such as the High Accuracy Radial velocity Planet Searcher \citep[HARPS;][]{HARPS} and the Ultraviolet and Visual Echelle Spectrograph \citep[UVES;][]{UVES}. These pioneering efforts helped establish a lower effective temperature ($T_{\rm eff}$) bound for cool-dwarf asteroseismology (see Fig.~\ref{fig:HRD}). However, limited RV precision and/or relatively small apertures restricted detections to the very brightest dwarfs.

Advances in Doppler precision, driven by the search for Earth-like planets, are providing a renewed opportunity for detecting low-amplitude oscillations, with Extremely Precise Radial Velocity (EPRV) techniques currently achieving $20$--$30\:{\rm cm\,s^{-1}}$ precision in a single exposure \citep[for a recent review, see][]{EPRV}. The few-minute oscillation periods typical of K dwarfs demand a fast observational cadence, requiring exposures as short as ${\sim} 25\:{\rm s}$ (with the remaining cadence budget allocated to readout). Achieving the necessary RV precision within such short integrations calls for an optimal combination of instrumental stability and large collecting area. The last couple of years have witnessed a groundbreaking resurgence of observational campaigns targeting K dwarfs (see Fig.~\ref{fig:HRD}), fueled by EPRV instruments mounted on large-aperture telescopes, namely, the Echelle SPectrograph for Rocky Exoplanets and Stable Spectroscopic Observations \citep[ESPRESSO;][]{ESPRESSO} on the Very Large Telescope (VLT) and the Keck Planet Finder \citep[KPF;][]{KPF} on the Keck I telescope. ESPRESSO and KPF are among the very few instrumental setups capable of achieving this level of performance, with a coordinated all-sky effort to characterize cool seismic dwarfs currently underway, leveraging both facilities.

A 6-night pilot campaign conducted with ESPRESSO on the fourth-magnitude K5 dwarf $\epsilon$~Indi \citep{Campante24} demonstrated the potential of EPRVs for cool-dwarf asteroseismology, enabling detection of solar-like oscillations with a peak amplitude of $2.6\:{\rm cm\,s^{-1}}$ \citep[or a mere ${\sim} 16\%$ of the solar value; see also][]{Lundkvist24}. This followed an earlier claim of p-mode detection in the nearly ten times fainter K3 dwarf HD~40307 \citep{ESPRESSO}, a result that was subsequently ruled out \citep{Campante24}. \citet{Hon24} later detected oscillations in the K0~V star $\sigma$~Draconis using only a single night of observations with KPF. While 20-second cadence TESS photometry from 14 sectors also showed oscillations for this star, the KPF data revealed oscillation modes over a considerably broader frequency range due to their higher S/N. More recently, based on four nights of data collected with KPF, \citet{Li25} detected oscillations in the K3~V star HD~219134, extracting 25 mode frequencies with angular degree $\ell\!=\!0$--$3$, and determined its benchmark mass, radius, and age via asteroseismology.

These results conclusively show that precise asteroseismology of cool dwarfs is now possible down to at least the mid-K spectral type, effectively opening a new domain in observational asteroseismology. This development carries significant implications for stellar astrophysics. In particular, mode amplitudes are determined by a delicate balance between energy supply and mode damping, both directly connected to the turbulent velocity field associated with convection \citep{HoudekDupret15}. Measurement of oscillation modes in cool dwarfs will thus shed light on the dynamical coupling between pulsations and near-surface convection in a regime yet unexplored. Theoretical calculations have suggested a scaling of the amplitudes of solar-like oscillations in Doppler velocity of the form $\varv_{\rm osc}\!\propto\!(L/M)^s$ \citep[][]{JCDFrandsen83,KjeldsenBedding95}, where $L$ and $M$ denote the stellar luminosity and mass, and the exponent $s$ is expected to lie in the range $0.7$--$1.5$ \citep{Houdek99,Samadi05,Samadi07}. Recent observations suggest that, for K dwarfs, mode amplitudes favor the upper bound of the predicted range for the exponent $s$ \citep{Campante24, Hon24, Li25}, consistent with a transition toward a steeper scaling as one moves from the G- into the K-dwarf regime, occurring at $T_{\rm eff}\!\sim\!5300\:{\rm K}$ \citep[see also][where a negative ${\rm d}s/{\rm d}T_{\rm eff}$ gradient was measured, based on a large ensemble of main-sequence and subgiant {\it Kepler} stars]{Verner11}. This emerging picture underscores the need for a deeper investigation into the physical origin of mode-amplitude scaling in the K-dwarf regime. Moreover, characterization of this scaling is key to accurately estimating the asteroseismic yield of the upcoming PLAnetary Transits and Oscillations of stars \citep[PLATO;][]{Goupil24,PLATO,Chaplin26} mission, whose P1 and P2 (``bright''), as well as P5 (``statistical'') stellar samples extend down to spectral type K7.

Here, we report on the ongoing asteroseismic survey of bright (apparent visual magnitude $V\!<\!6$) southern and equatorial K dwarfs (declination $\delta\!<\!+30^\circ$) with ESPRESSO, presenting recent observational campaigns on \object{GJ~570~A} (K4~V) and \object{GJ~783~A} (K2.5~V). We are able to detect and measure solar-like oscillations in the EPRV data of both stars. These stars thus join a still sparse but growing sample of seismic K dwarfs (see Fig.~\ref{fig:HRD}), enhancing the statistical basis for studies of convection--pulsation coupling. The remainder of the paper is organized as follows. Section~\ref{sect:obs} describes the ESPRESSO observations of GJ~570~A and GJ~783~A and the data reduction procedure used to obtain the RV time series. The spectroscopic determination of the stellar atmospheric parameters and chromospheric emission is presented in Sect.~\ref{sect:spec}. A detailed asteroseismic analysis follows in Sect.~\ref{sect:asteroseismology}. In Sect.~\ref{sect:amp}, we place the measured oscillation amplitudes of GJ~570~A and GJ~783~A in a broader context and highlight the tension between the empirical mode-amplitude scaling for K dwarfs and predictions from 3D convection simulations. Section~\ref{sect:discussion} summarizes and discusses the results.

\section{Observations and data reduction}\label{sect:obs}

GJ~570~A and GJ~783~A are nearby ($d\!\sim\!6\:{\rm pc}$), fifth-magnitude dwarfs of spectral type K4 and K2.5, respectively. GJ~570~A (HD~131977, HIP~73184, HR~5568, TIC~287157634) is the primary component of a quadruple system that further comprises an M1.5~V--M3~V close binary and a widely separated late-T brown dwarf \citep[][]{Burgasser00}. It is a BY~Draconis-type variable, characterized by photometric variability due to rotational modulation by starspots and other surface active regions, with variable star designation KX~Librae \citep[][]{GCVS}. A rotation period of ${\sim} 40\:{\rm d}$ has been reported for the star \citep{Cumming99,Geballe01}. GJ~783~A (HD~191408, HIP~99461, HR~7703, TIC~389198736) is a metal-poor star and the primary in a wide binary system with an M4~V companion \citep[][]{Allen00}. Table~\ref{tab:param} compiles the stellar properties of GJ~570~A and GJ~783~A.

GJ~570~A and GJ~783~A are both part of the ExEP HWO Precursor Science Stars list \citep[][]{HWO} compiled by the NASA Exoplanet Exploration Program (ExEP), being among the best targets for a systematic imaging survey of habitable zones by the future Habitable Worlds Observatory (HWO). Both stars have been observed by TESS in 20-second cadence. GJ~783~A was included in the TESS Luminaries sample \citep{Lund25}, based on Sector~67 data, and we have since reprocessed the data in the same manner to include Sectors~94 and 105. Similarly, GJ~570~A was processed following \citet{Lund25}, based on Sector~91 data. No clear detection of solar-like oscillations is obtained for either star in the TESS photometry. GJ~570~A will be observed again by TESS in Sector~115.

\begin{table}[t]
\small
\caption{Stellar properties of GJ~570~A and GJ~783~A.}
\label{tab:param}
\centering
\renewcommand{\tabcolsep}{0.8mm}
\renewcommand{\arraystretch}{1.15}
\begin{tabular}{l | c @{\hspace{-2.0mm}} c | c @{\hspace{-2.0mm}} c @{\hspace{-0.4mm}}}
\hline\hline
\noalign{\smallskip}
\multicolumn{1}{l}{} & \multicolumn{2}{c}{\textbf{GJ~570~A}} & \multicolumn{2}{c}{\textbf{GJ~783~A}} \\
\noalign{\smallskip}
\hline
\noalign{\smallskip}
Parameter & Value & Source\phantom{\;} & Value & Source\phantom{\;} \\
\noalign{\smallskip}
\hline
\noalign{\smallskip}
\multicolumn{5}{c}{Basic data} \\
\noalign{\smallskip}
\hline
\noalign{\smallskip}
Sp.~Type & K4~V & 1 & K2.5~V & 1 \\
$V$ (Johnson) & 5.72\,\tablefootmark{a} & 2,\,3 & 5.32 & 3 \\
$B\!-\!V$ (Johnson) & 1.02 & 3 & 0.87 & 3 \\
$\delta$ ($^\circ$) & $-21.42$ & 4 & $-36.11$ & 4 \\
$\varv_{\rm rad}$ (${\rm km\,s^{-1}}$) & $26.75\pm0.12$ & 4 & $-129.27\pm0.12$ & 4 \\
$d$ (pc)\,\tablefootmark{b} & $5.885^{+0.002}_{-0.002}$ & 5 & $6.011^{+0.004}_{-0.003}$ & 5 \\
\noalign{\smallskip}
\hline
\noalign{\smallskip}
\multicolumn{5}{c}{Spectroscopy} \\
\noalign{\smallskip}
\hline
\noalign{\smallskip}
$T_{\rm eff}$ (K) & $4629\pm123$ & 6 & $4854\pm83$ & 6 \\
$[{\rm Fe}/{\rm H}]$ (dex) & $-0.04\pm0.07$ & 6 & $-0.63\pm0.06$ & 6 \\
$[\alpha/{\rm Fe}]$ (dex) & $0.04\pm0.07$ & 6 & $0.22\pm0.06$ & 6 \\
$\log g$ (dex) & $4.54\pm0.07$ & 6 & $4.49\pm0.05$ & 6 \\
$\log R^{\prime}_{\rm HK}$ (dex) & $-4.648\pm0.009$ & 6 & $-5.034\pm0.003$ & 6 \\
\noalign{\smallskip}
\hline
\noalign{\smallskip}
\multicolumn{5}{c}{Interferometry} \\
\noalign{\smallskip}
\hline
\noalign{\smallskip}
$R$ ($R_\odot$) & $0.715\pm0.009$ & 7 & $\cdots$ & $\cdots$ \\
\noalign{\smallskip}
\hline
\noalign{\smallskip}
\multicolumn{5}{c}{Asteroseismology} \\
\noalign{\smallskip}
\hline
\noalign{\smallskip}
$\Delta\nu$ ($\mu$Hz) & $197.29\pm0.22$ & 6 & $191.48\pm0.04$ & 6 \\
$\delta\nu_{02}$ ($\mu$Hz) & $11.64\pm0.73$ & 6 & $12.48\pm0.32$ & 6 \\
$\varepsilon$ & $1.328\pm0.027$ & 6 & $1.517\pm0.004$ & 6 \\
$\nu_{\rm max}$ ($\mu$Hz) & $4705\pm105$ & 6 & $4455\pm70$ & 6 \\
$\varv_{\rm osc,max}$ (${\rm cm\,s^{-1}}$) & $2.57\pm0.52$ & 6 & $4.90\pm0.81$ & 6 \\
\noalign{\smallskip}
\hline
\noalign{\smallskip}
\multicolumn{5}{c}{Frequency modeling\,\tablefootmark{c}} \\
\noalign{\smallskip}
\hline
\noalign{\smallskip}
$M$ ($M_\odot$) & $0.679\!\pm\!0.017\!\pm\!0.012$ & 6 & $0.686\!\pm\!0.012\!\pm\!0.005$ & 6 \\
$R$ ($R_\odot$) & $0.685\!\pm\!0.006\!\pm\!0.004$ & 6 & $0.697\!\pm\!0.005\!\pm\!0.001$ & 6 \\
$t_{\rm age}$ (Gyr) & $8.9\!\pm\!2.2\!\pm\!1.1$ & 6 & $9.0\!\pm\!0.6\!\pm\!0.7$ & 6 \\
\noalign{\smallskip}
\hline
\end{tabular}
\tablefoot{
\tablefoottext{a}{Magnitude at maximum brightness (BY~Draconis-type variable).}
\tablefoottext{b}{Median of the geometric distance posterior.}
\tablefoottext{c}{The quoted uncertainties are, respectively, the statistical and systematic errors.}
}
\tablebib{
(1)~\citet{NStars};
(2)~\citet{GCVS};
(3)~\citet{HIP};
(4)~\citet{GaiaDR3};
(5)~\citet{Bailer-Jones21};
(6)~This work;
(7)~\citet{Rains20}.
}
\end{table}

We observed GJ~570~A (GJ~783~A) for five (six) consecutive nights with ESPRESSO in May (September) 2024. Observations were carried out in single-Unit Telescope (single-UT) high-resolution mode ($1\!\times\!1$ binning, fast readout; median resolving power $R\!\approx\!138{,}000$). The GJ~570~A campaign was initially scheduled for six nights, but the first night and the latter half of the fourth (or third effective) night were lost to high winds and overcast skies, during which the telescope remained closed, resulting in five usable nights of data. The remainder of the campaign was affected by extended periods of thin and thick cirrus, as well as occasional strong winds. In contrast, weather conditions during the GJ~783~A campaign were considerably more favorable, with photometric or mostly clear skies throughout. The average nightly seeing\footnote{Normalized to an airmass of one and to a wavelength of $500\:{\rm nm}$ ($V$ band).} varied from $0\farcs59$--$1\farcs28$ ($0\farcs53$--$0\farcs86$) for GJ~570~A (GJ~783~A).

\begin{figure*}[!t]
\centering
\includegraphics[width=0.495\textwidth,trim=1.5cm 0.75cm 0.25cm 0cm,clip]{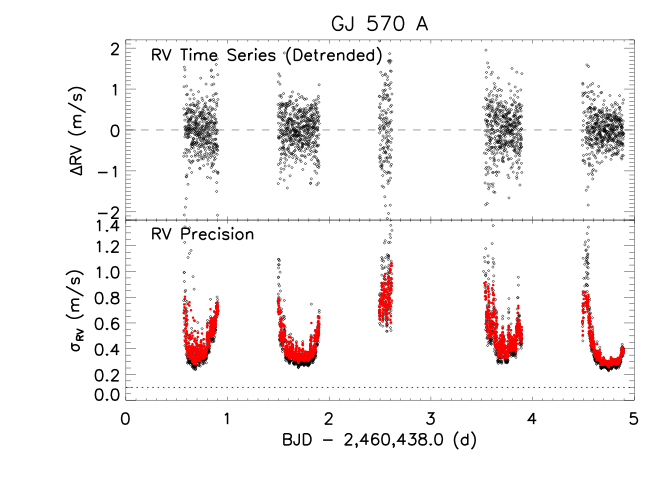}
\includegraphics[width=0.495\textwidth,trim=1.5cm 0.75cm 0.25cm 0cm,clip]{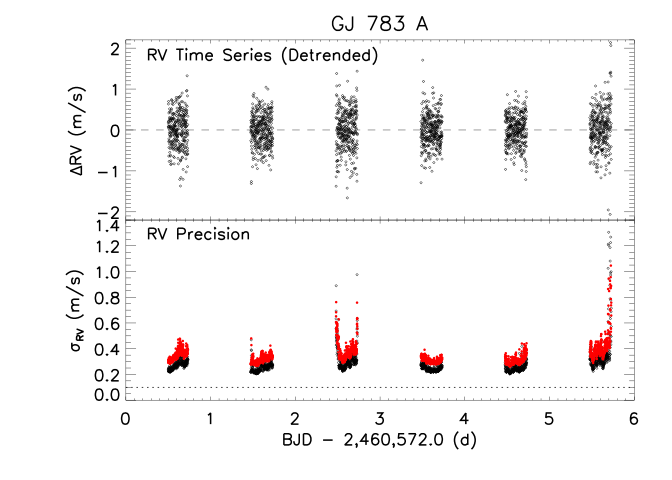}
\caption{Radial-velocity time series from ESPRESSO observations of GJ~570~A (left) and GJ~783~A (right). \emph{Upper subpanels:} Detrended RV time series (see text for details). \emph{Lower subpanels:} Internal RV precision as returned by the ESPRESSO DRS (open black diamonds) and recalibrated measurement uncertainties (filled red circles; see Sect.~\ref{sect:freq}). Horizontal dotted lines mark the instrumental noise floor of $10\:{\rm cm\,s^{-1}}$ quoted by \citet{ESPRESSO}.}
\label{fig:ts}
\end{figure*}

We obtained $N\!=\!2369$ (2133) spectra\footnote{After discarding spectra for which the atmospheric dispersion correction failed, we were left with $N\!=\!2250$ (2125) usable RV data points for GJ~570~A (GJ~783~A).} for GJ~570~A (GJ~783~A) using a fixed exposure time of $25\:{\rm s}$ and a median cadence of $\Delta t\!\sim\!61\:{\rm s}$ between exposures, corresponding to a Nyquist frequency of $\nu_{\rm Nyq}\!=\!1/(2\Delta t)\!\sim\!8.2\:{\rm mHz}$. The spectra were reduced using the ESPRESSO Data Reduction Software \citep[DRS;][]{ESPRESSO} and RV time series\footnote{Raw RV time series can be downloaded for both targets from \href{https://doi.org/10.5281/zenodo.17236515}{this Zenodo repository}.} were derived via cross-correlation, similarly to \citet[][]{Campante24}. The resulting RV time series are displayed in the upper subpanels of Fig.~\ref{fig:ts}. These have been detrended by applying a high-pass filter to the raw time series, performed night by night using a triangular smoothing function with a 1.2-hour cutoff. This was done to remove slow modulations and intranight trends, thereby isolating the high-frequency signals of interest. The lower subpanels of Fig.~\ref{fig:ts} show the photon-noise-limited RV precision per data point, $\sigma_{\rm RV}$, as returned by the ESPRESSO DRS (open black diamonds), with average values of $\langle\sigma_{\rm RV}\rangle\!=\!46\:{\rm cm\,s^{-1}}$ for GJ~570~A and $29\:{\rm cm\,s^{-1}}$ for GJ~783~A. The corresponding root-mean-square (rms) scatter of the detrended RV time series is $54\:{\rm cm\,s^{-1}}$ ($45\:{\rm cm\,s^{-1}}$), exceeding the photon noise in both cases. This excess is attributed to the presence of solar-like oscillations as well as residual instrumental noise.

\section{Spectroscopy}\label{sect:spec}

\subsection{Stellar atmospheric parameters}\label{sect:atmo}
Estimation of the effective temperature, $T_{\rm eff}$, surface gravity, $\log g$, and metallicity, $[{\rm Fe}/{\rm H}]$, was performed using the \texttt{ARES}~+~\texttt{MOOG} method, described in detail in \citet[][]{Santos13,Sousa14,Sousa21}. We used the Automatic Routine for line Equivalent widths in stellar Spectra\footnote{\url{https://github.com/sousasag/ARES}} \citep[\texttt{ARES};][]{Sousa07,Sousa15} to measure the equivalent widths of lines in co-added ESPRESSO spectra. Since both stars are cooler than $5200\:{\rm K}$, we adopted the line list introduced by \citet[][]{Tsantaki13}, which is well suited for K dwarfs. The best-fit spectroscopic parameters were determined by employing a minimization procedure to enforce ionization and excitation equilibrium. This procedure used a grid of Kurucz model atmospheres \citep{Kurucz93} and the latest version of the \texttt{MOOG} \citep{Sneden73} radiative transfer code. Moreover, we followed the procedure described in \citet[][]{Sousa21} to derive a trigonometric $\log g$ by combining the spectroscopic $T_{\rm eff}$ with {\it Gaia} parallax and photometry \citep{GaiaDR3}, which provides a more accurate estimate than the purely spectroscopic $\log g$. We list the trigonometric $\log g$ in Table~\ref{tab:param}, along with $T_{\rm eff}$ and $[{\rm Fe}/{\rm H}]$.

Based on the derived atmospheric parameters, we determined the abundances of Mg, Si, and Ti using a classical curve-of-growth analysis \citep[e.g.,][]{Adibekyan12,Adibekyan15}. As in the stellar parameter determination, we used \texttt{ARES} to measure the equivalent widths of the spectral lines of these elements, and employed a grid of Kurucz model atmospheres together with \texttt{MOOG} to convert the equivalent widths into abundances under the assumption of local thermodynamic equilibrium (LTE). Finally, we estimated the $\alpha$-element to iron abundance ratio, $[\alpha/{\rm Fe}]$, as a weighted mean of the individual Mg, Si, and Ti abundances, using the inverse squared abundance uncertainties as weights (see Table~\ref{tab:param}).

For completeness, we also applied the same procedure to high-quality ESPRESSO spectra of $\epsilon$~Indi, which had been previously analyzed by our team in \citet{Campante24}, but without a homogeneous derivation of its atmospheric parameters. Using the methodology described above, we determined $T_{\rm eff}\!=\!4581\pm130\:{\rm K}$, $\log g\!=\!4.51\pm0.08\:{\rm dex}$, $[{\rm Fe}/{\rm H}]\!=\!-0.18\pm0.07\:{\rm dex}$, and $[\alpha/{\rm Fe}]\!=\!0.00\pm0.08\:{\rm dex}$ for this star.

\subsection{Chromospheric emission}\label{sect:chromo}
To estimate the mean activity level and investigate possible long-term trends, we supplemented the single-epoch ESPRESSO observations with archival HARPS data \citep[][]{GomesdaSilva21}. The HARPS data span from 2004 to 2012 for GJ~570~A, and from 2003 to 2008 for GJ~783~A. We extracted the S-index from the \ion{Ca}{II} H\&K lines \citep{Duncan91} using \texttt{ACTIN 2}\footnote{\url{https://github.com/gomesdasilva/ACTIN2}} \citep{GomesdaSilva18,GomesdaSilva21}, and computed the chromospheric emission ratio, $\log R^{\prime}_{\rm HK}$, employing the \texttt{pyrhk}\footnote{\url{https://github.com/gomesdasilva/pyrhk}} package and the methodology outlined in \citet{GomesdaSilva21}. For each instrument, we performed an iterative $3\sigma$ clipping on the $\log R^{\prime}_{\rm HK}$ time series to remove outliers before binning the data. The retained measurements were binned into 26 (HARPS) and 5 (ESPRESSO) observing nights for GJ~570~A, and 22 (HARPS) and 6 (ESPRESSO) nights for GJ~783~A. We searched for significant periodicities in the HARPS time series of both stars using a periodogram analysis \citep{GLS}, but found no signals with false-alarm probabilities below $10\%$.

\begin{figure*}[!t]
\centering
\includegraphics[width=0.495\textwidth,trim=1.5cm 0.75cm 0.25cm 0cm,clip]{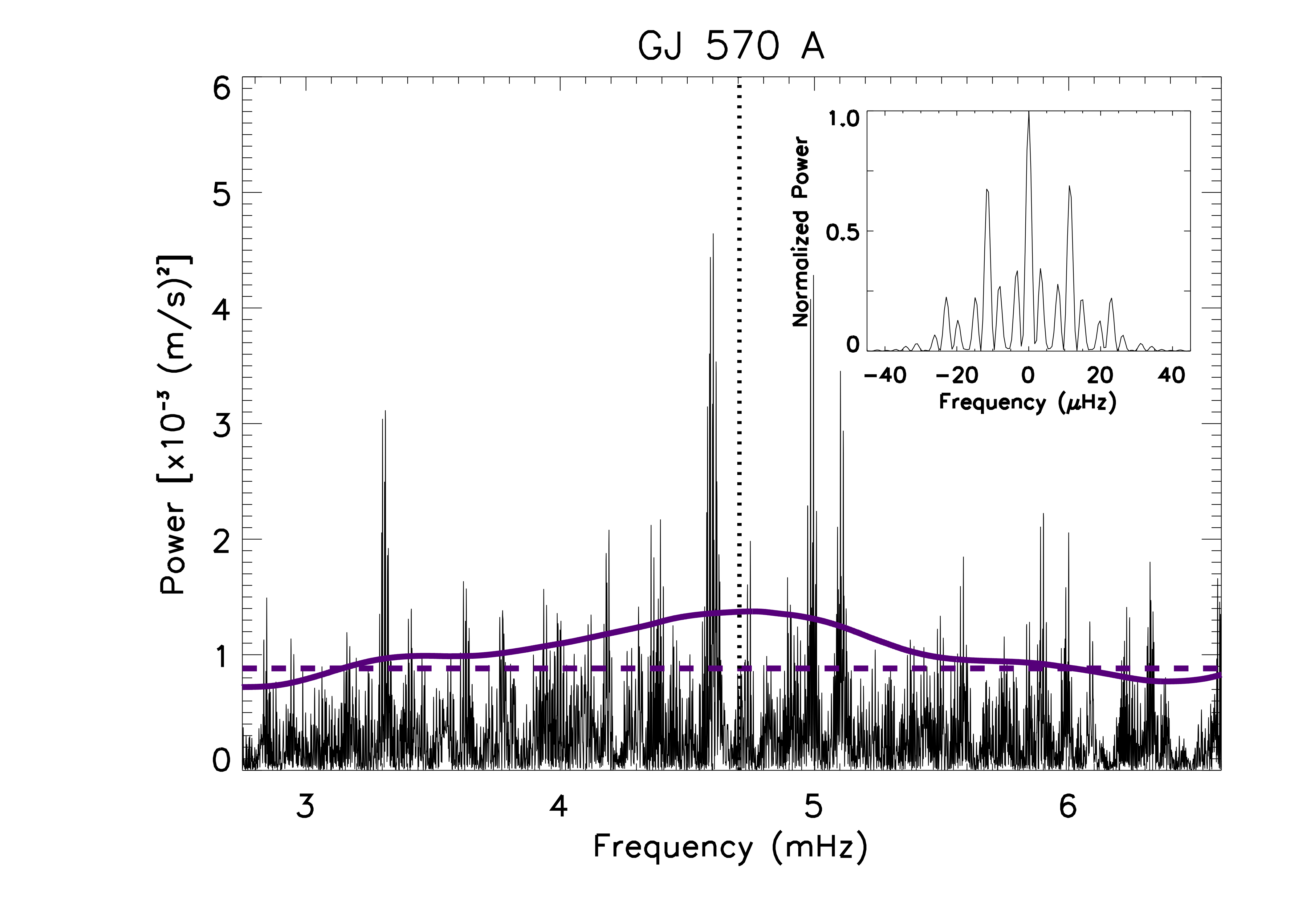}
\includegraphics[width=0.495\textwidth,trim=1.5cm 0.75cm 0.25cm 0cm,clip]{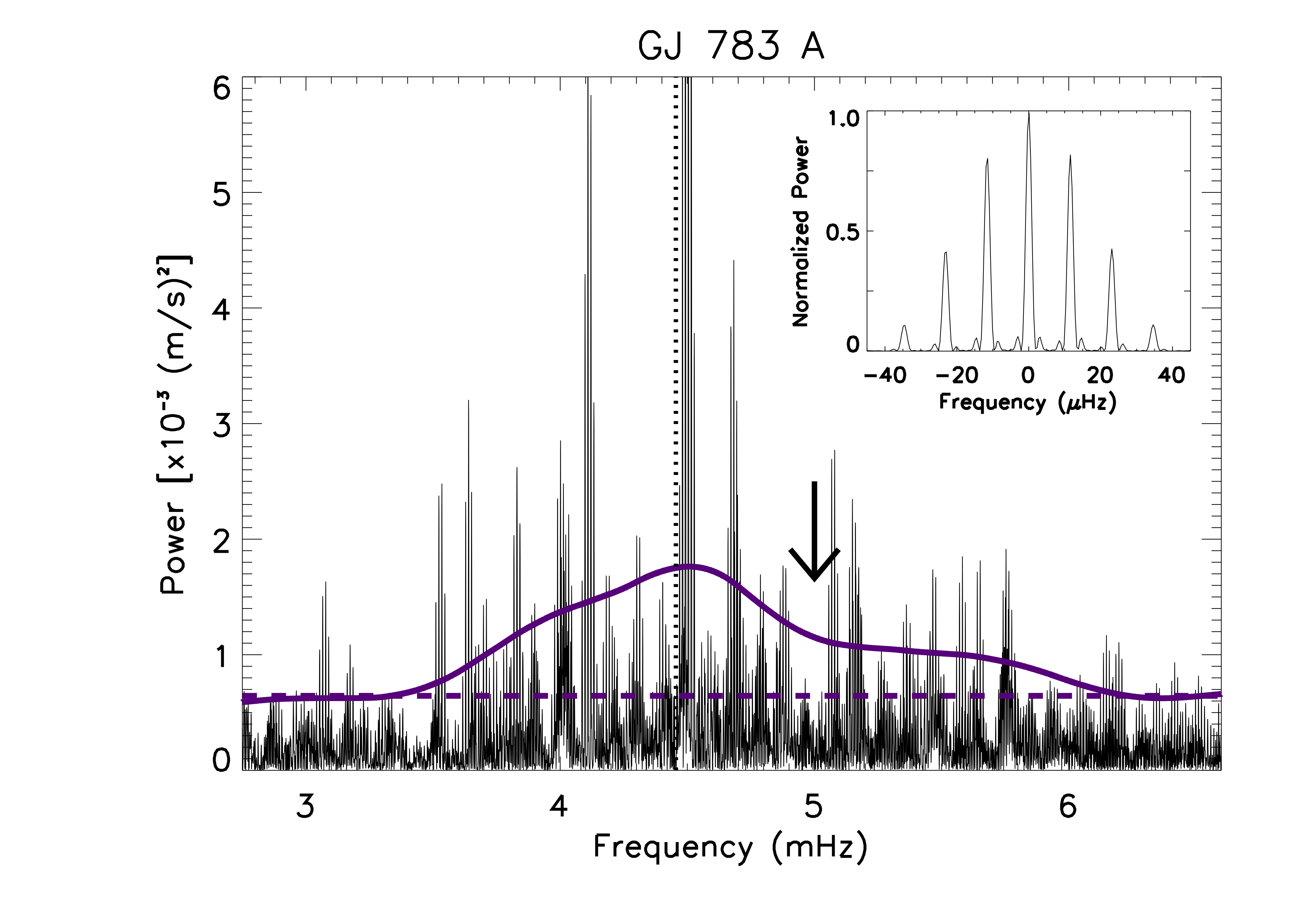}
\caption{Power spectra of GJ~570~A (left) and GJ~783~A (right), shown oversampled for clarity. Excess power from solar-like oscillations is visible in both panels, centered slightly below $5\:{\rm mHz}$ (vertical dotted lines mark the estimated $\nu_{\rm max}$). Smoothed power spectra (solid curves) and the corresponding white-noise levels (horizontal dashed lines), both scaled by a factor of three, are overplotted to highlight this excess. The smoothed curves are intended for visualization and differ from the heavily smoothed power spectra computed in Sect.~\ref{sect:scaling}. The arrow marks an apparent dip in the oscillation power of GJ~783~A (see Sect.~\ref{sect:sim}). Insets display the spectral windows, showing prominent sidelobes at multiples of $\pm11.57\:{\rm \mu Hz}$, or $\pm1\:{\rm d^{-1}}$, resulting from the single-site nature of the observations.}
\label{fig:ps}
\end{figure*}

We next evaluated the mean activity levels, computing the weighted mean of $\log R^{\prime}_{\rm HK}$ for each instrument separately. The associated uncertainties were estimated as the standard errors on the mean. For GJ~570~A, the chromospheric emission ratios derived from HARPS and ESPRESSO are nearly identical, with $\log R^{\prime}_{\rm HK}\!=\!-4.649\pm0.005\:{\rm dex}$ and $-4.648\pm0.009\:{\rm dex}$, respectively. This consistency suggests that GJ~570~A may have maintained a stable activity level across the two-decade interval between 2004 and 2024. In contrast, for GJ~783~A, the HARPS data yield $\log R^{\prime}_{\rm HK}\!=\!-4.983\pm0.003\:{\rm dex}$, whereas the more recent ESPRESSO observations indicate a lower activity level of $\log R^{\prime}_{\rm HK}\!=\!-5.034\pm0.003\:{\rm dex}$, pointing to a possible long-term decline in chromospheric activity between the 2003--2008 and 2024 epochs. Overall, these activity levels indicate that GJ~570~A is a moderately active star, lying close to the Vaughan--Preston gap \citep{VaughanPreston80}, while GJ~783~A resides near the peak of the inactive-star distribution \citep{Henry96}.

\section{Asteroseismology}\label{sect:asteroseismology}

\subsection{Oscillation frequencies}\label{sect:freq}
Power spectra of the detrended RV time series were computed based on the Lomb--Scargle periodogram \citep[][]{Lomb76,Scargle82}, with measurement uncertainties adopted as statistical weights, i.e., $w_i\!=\!1/\sigma^2_{{\rm RV},i}$ \citep[][]{Frandsen95}, where the subscript $i$ denotes individual measurements. To further optimize the noise floor in the power spectra of both stars, these weights were adjusted to account for small fractions of bad data points ($1.2\%$ and $0.9\%$ of the data points were removed for GJ~570~A and GJ~783~A, respectively) and for night-to-night variations in the noise level, following a well-established procedure \citep[e.g.,][]{Bedding04,Arentoft08,Kjeldsen25}.

The resulting noise-optimized power spectra are shown in Fig.~\ref{fig:ps}, revealing excess power above the white-noise level from solar-like oscillations centered at frequencies slightly below $5\:{\rm mHz}$ (corresponding to typical oscillation periods of 3--4 minutes), although with a noticeably lower S/N for GJ~570~A (left panel). We estimated the frequency of maximum oscillation amplitude, $\nu_{\rm max}$, as the frequency corresponding to the peak of the oscillation envelope obtained after heavily smoothing the power spectrum with a Gaussian kernel and subtracting the white-noise background. The full procedure used to convert the smoothed power spectra into amplitudes per radial mode is described in Sect.~\ref{sect:scaling}. We obtained $\nu_{\rm max}\!=\!4705\pm105\:{\rm \mu Hz}$ for GJ~570~A and $4455\pm70\:{\rm \mu Hz}$ for GJ~783~A. An apparent dip in the oscillation power of the higher-S/N star GJ~783~A is also visible in the right panel of Fig.~\ref{fig:ps} above $\nu_{\rm max}$. This feature is reminiscent of similar apparent local minima reported in the oscillation spectra of the K dwarfs $\alpha$~Centauri~B \citep{Kjeldsen05} and HD~219134 \citep{Li25}. We return to a discussion of this feature in Sect.~\ref{sect:sim}.

The white-noise level in the amplitude spectrum---defined as the square root of the power spectrum---was measured at high frequencies ($6$--$8\:{\rm mHz}$), yielding $\sigma_{\rm white}\!=\!1.51\:{\rm cm\,s^{-1}}$ for GJ~570~A and $\sigma_{\rm white}\!=\!1.29\:{\rm cm\,s^{-1}}$ for GJ~783~A. For comparison, the white-noise level reported for the ESPRESSO observations of $\epsilon$~Indi ($V\!=\!4.69$) is $\sigma_{\rm white}\!=\!0.94\:{\rm cm\,s^{-1}}$ \citep[][]{Campante24}, consistent with its higher apparent brightness, while the number of data points, $N$, is very similar for all three targets. The recalibrated measurement uncertainties shown in the lower subpanels of Fig.~\ref{fig:ts} (filled red circles) reflect the adjusted weights described above and were further scaled to reproduce the high-frequency noise level in the amplitude spectrum \citep{Butler04}.

The observed oscillation modes are acoustic (p) modes of high radial order, $n$, and low angular degree, $\ell$, with their frequencies well approximated by the asymptotic relation \citep[][]{Tassoul80}:
\begin{equation}
\nu_{n,\ell} \simeq \Delta\nu \left(n+\frac{\ell}{2}+\varepsilon\right) - \delta\nu_{0\ell} \, ,
\label{eq:asymptotic}
\end{equation}
where $\Delta\nu$ is the large frequency separation between modes of the same degree and consecutive order, scaling as the square root of the mean stellar density; the small frequency separations $\delta\nu_{0\ell}$ (e.g., $\delta\nu_{02}$) are measured between modes of different degree and adjacent orders, being sensitive to the sound-speed gradient near the core in main-sequence stars; and the dimensionless offset, $\varepsilon$, is determined by the reflection properties of the surface layers.

\begin{figure}[!t]
\centering
\includegraphics[width=0.46\textwidth,trim=0cm 0cm 0cm 0cm,clip]{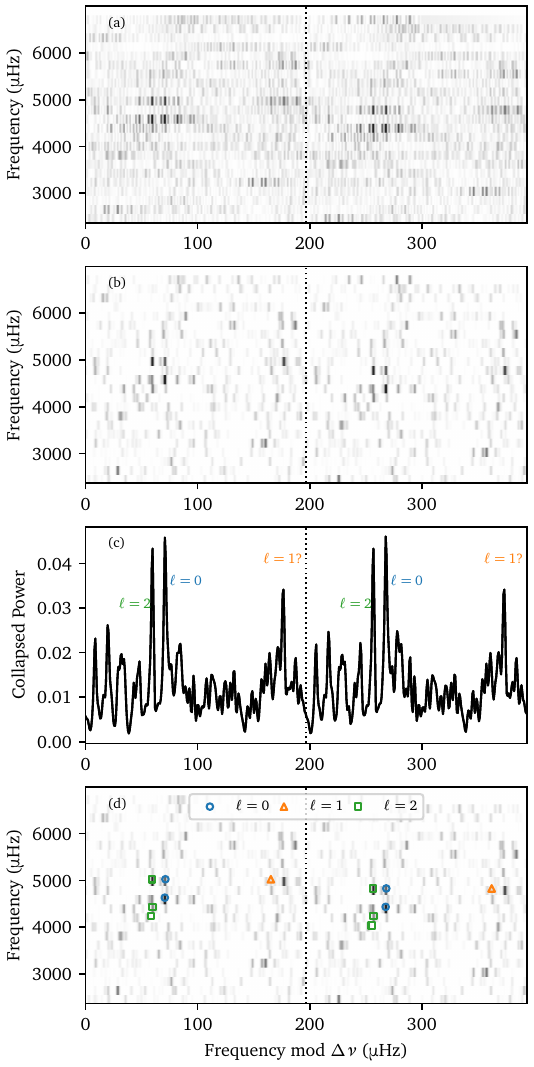}
\caption{Mode identification and extraction for GJ~570~A (see text for details). \emph{Panel (a):} Replicated \'echelle diagram of the observed power spectrum. \emph{Panel (b):} Replicated \'echelle diagram of the deconvolved power spectrum. \emph{Panel (c):} Collapsed (vertically) deconvolved \'echelle diagram, with power ridges labeled by their corresponding angular degree. \emph{Panel (d):} Same as panel (b), but highlighting the extracted modes. Vertical dotted lines indicate the estimated $\Delta\nu$. A slight offset is seen between the only extracted $\ell\!=\!1$ mode and the tallest peak in the corresponding power ridge in panel (c), consistent with the marginal S/N of this star (${\rm S/N}\!\approx\!2.5$--$3$).}
\label{fig:echelle_GJ570A}
\end{figure}

\begin{figure}[!t]
\centering
\includegraphics[width=0.46\textwidth,trim=0cm 0cm 0cm 0cm,clip]{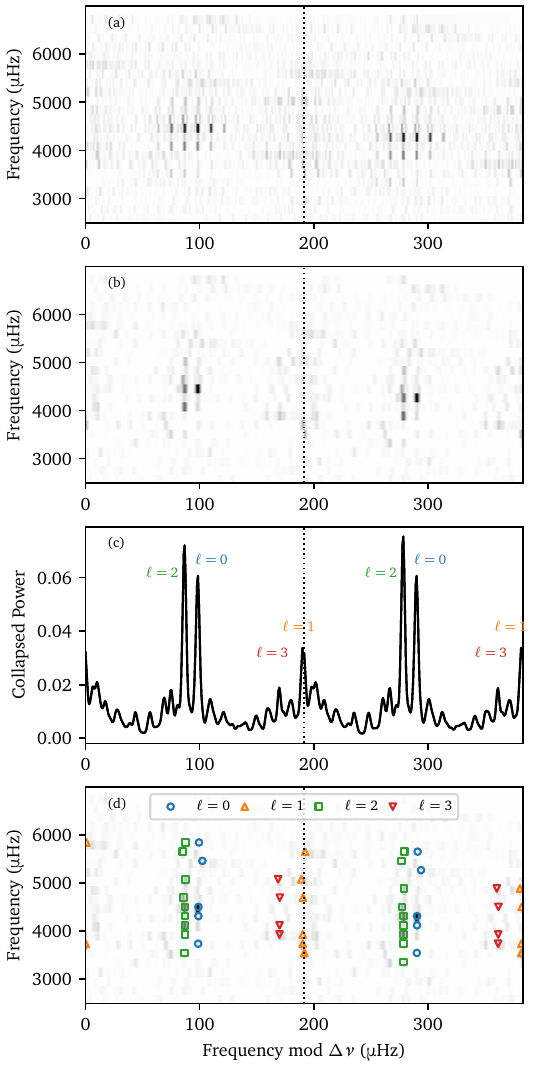}
\caption{Mode identification and extraction for GJ~783~A (see text for details). \emph{Panels (a)--(d):} Replicated observed and deconvolved \'echelle diagrams, collapsed power, and extracted modes, respectively, with vertical dotted lines indicating the estimated $\Delta\nu$. The panels are arranged in the same format as in Fig.~\ref{fig:echelle_GJ570A}.}
\label{fig:echelle_GJ783A}
\end{figure}

The presence of daily aliases in the power spectra of both stars (see corresponding spectral windows in the insets of Fig.~\ref{fig:ps}), appearing at multiples of $\pm11.57\:{\rm \mu Hz}$, or $\pm1\:{\rm d^{-1}}$, around genuine peaks, greatly complicates the process of mode identification and extraction. Measurement of the $\delta\nu_{02}$ spacing is particularly challenging, as it is expected to take values very close to $1\:{\rm d^{-1}}$ for K dwarfs, based on stellar models \citep[][]{White11_1,White11_2}. To recover the true power spectra, we therefore deconvolved the observed power spectra using their spectral windows by means of the Gold deconvolution algorithm \citep[][]{Gold64}, as implemented by \citet{Morhac03}. A similar deconvolution strategy was employed by \citet{Li25} in the asteroseismic analysis of HD~219134. Figures~\ref{fig:echelle_GJ570A}(a),~(b) and \ref{fig:echelle_GJ783A}(a),~(b) show the \'echelle diagrams of GJ~570~A and GJ~783~A, respectively, before and after deconvolution, revealing $\ell\!=\!0$--$2$ power ridges (as well as $\ell\!=\!3$ in the case of GJ~783~A). These power ridges are made even more apparent in Figs.~\ref{fig:echelle_GJ570A}(c) and \ref{fig:echelle_GJ783A}(c), which display the collapsed (along the vertical direction) deconvolved \'echelle diagrams, thereby facilitating mode identification and enabling the estimation of guess frequencies for mode extraction. For GJ~783~A, the collapsed spectrum shows a lower-than-expected power ratio between the $\ell\!=\!1$ and $\ell\!=\!0$ ridges, given the relative visibility expected for stellar velocity observations \citep[e.g.,][]{Kjeldsen08}. However, inspection of the individual radial orders reveals substantial variations in the relative $\ell\!=\!1$ and $\ell\!=\!0$ amplitudes across the frequency range, rather than a systematic suppression of the $\ell\!=\!1$ modes. We therefore interpret the low relative $\ell\!=\!1$ power in the collapsed spectrum as being consistent with stochastic excitation over the finite observing interval.

We extracted the mode frequencies by simultaneously fitting sine waves at these guess frequencies to the RV time series, retaining only those modes with amplitudes exceeding 2.5 times the local noise level in the amplitude spectrum, $\sigma_{\rm amp}(\nu)$ (i.e., ${\rm S/N}\!\ge\!2.5$; see Figs.~\ref{fig:echelle_GJ570A}(d) and \ref{fig:echelle_GJ783A}(d)). Given the short duration of the RV time series, the modes are only marginally resolved, which justifies the use of coherent sine-wave fits. This procedure resulted in a total of 6 and 24 modes being extracted for GJ~570~A and GJ~783~A, respectively, as listed in Tables~\ref{tab:freq_GJ570A} and \ref{tab:freq_GJ783A}. The listed frequencies were corrected for the stellar line-of-sight Doppler velocity shift following \citet{Davies14}, using the corresponding {\it Gaia} radial velocities, $\varv_{\rm rad}$ \citep{GaiaDR3}. Frequency uncertainties were estimated using the analytical expression \citep{ToutainAppourchaux94,KjeldsenBedding12}:
\begin{equation}
    \sigma\,(\nu) \approx 0.44 \, \sigma_{\rm amp}(\nu) \, a^{-1} \sqrt{T_{\rm obs}^{-2} + \tau^{-2}} \, ,
\label{eq:errfreq}
\end{equation}
where $a$ is the mode amplitude, $T_{\rm obs}$ is the effective observing duration, and $\tau$ is the mode lifetime. In the uncertainty calculation, we assumed a conservative mode lifetime of 3 days \citep[cf.][]{Campante24}. In principle, the mode lifetime can be constrained observationally by measuring the scatter of the radial-mode frequencies about their power ridge in the \'echelle diagram and comparing it with simulations \citep[e.g.,][]{Bedding04,Kjeldsen05}. We attempted such an analysis for the higher-S/N star GJ~783~A; however, the measured scatter is too small to yield a robust lower limit on $\tau$. Finally, we fitted Eq.~(\ref{eq:asymptotic}) to both frequency sets to estimate the asymptotic parameters $\Delta\nu$, $\delta\nu_{02}$, and $\varepsilon$ (see Table~\ref{tab:param}). Owing to the limited number of extracted modes for the lower-S/N star GJ~570~A, the corresponding $\delta\nu_{02}$ and $\varepsilon$ values should be regarded as tentative, though they remain broadly consistent with stellar model predictions \citep[cf.][]{Campante24}; note that a small offset in $\varepsilon$ relative to the models is expected due to the surface effect \citep[e.g.,][]{Kjeldsen08_surf_corr,BallGizon14}.

\begin{table}[!t]
\small
\caption{Oscillation frequencies of GJ~570~A.}
\label{tab:freq_GJ570A}
\centering
\renewcommand{\arraystretch}{1.15}
\begin{tabular}{c c c c c}
\hline\hline
\noalign{\smallskip}
$n$ & $\ell$ & $\nu_{n,\ell}$ (${\rm \mu Hz}$) & $\sigma(\nu_{n,\ell})$ (${\rm \mu Hz}$) & ${\rm S/N}$ \\
\noalign{\smallskip}
\hline
\noalign{\smallskip}
22 & 0 & 4602.38 & 0.73 & 2.8 \\
24 & 0 & 4996.98 & 0.75 & 2.7 \\
24 & 1 & 5091.01 & 0.65 & 3.1 \\
19 & 2 & 4195.93 & 0.76 & 2.6 \\
20 & 2 & 4394.34 & 0.78 & 2.6 \\
23 & 2 & 4985.14 & 0.71 & 2.8 \\
\noalign{\smallskip}
\hline
\end{tabular}
\tablefoot{Listed frequencies were corrected for the stellar line-of-sight Doppler velocity shift, which for this star amounts to approximately $+0.4\:{\rm \mu Hz}$ at $\nu_{\rm max}$.}
\end{table}

\begin{table}[!t]
\small
\caption{Oscillation frequencies of GJ~783~A.}
\label{tab:freq_GJ783A}
\centering
\renewcommand{\arraystretch}{1.15}
\begin{tabular}{c c c c c}
\hline\hline
\noalign{\smallskip}
$n$ & $\ell$ & $\nu_{n,\ell}$ (${\rm \mu Hz}$) & $\sigma(\nu_{n,\ell})$ (${\rm \mu Hz}$) & ${\rm S/N}$ \\
\noalign{\smallskip}
\hline
\noalign{\smallskip}
18 & 0 & 3736.72 & 0.72 & 2.7 \\
21 & 0 & 4311.34 & 0.57 & 3.4 \\
22 & 0 & 4502.88 & 0.52 & 3.7 \\
27 & 0 & 5463.70 & 0.74 & 2.6 \\
29 & 0 & 5843.65 & 0.56 & 3.5 \\
17 & 1 & 3638.21 & 0.50 & 3.9 \\
18 & 1 & 3828.19 & 0.45 & 4.3 \\
19 & 1 & 4019.73 & 0.52 & 3.7 \\
23 & 1 & 4785.89 & 0.64 & 3.0 \\
25 & 1 & 5167.40 & 0.69 & 2.8 \\
28 & 1 & 5745.15 & 0.66 & 2.9 \\
16 & 2 & 3533.45 & 0.52 & 3.7 \\
18 & 2 & 3916.53 & 0.70 & 2.8 \\
19 & 2 & 4108.07 & 0.36 & 5.4 \\
20 & 2 & 4299.61 & 0.65 & 3.0 \\
21 & 2 & 4491.15 & 0.41 & 4.7 \\
22 & 2 & 4681.13 & 0.44 & 4.4 \\
24 & 2 & 5065.77 & 0.64 & 3.0 \\
27 & 2 & 5638.04 & 0.77 & 2.5 \\
28 & 2 & 5831.92 & 0.53 & 3.6 \\
18 & 3 & 3999.40 & 0.46 & 4.2 \\
19 & 3 & 4190.94 & 0.63 & 3.1 \\
22 & 3 & 4765.56 & 0.72 & 2.7 \\
24 & 3 & 5147.07 & 0.72 & 2.7 \\
\noalign{\smallskip}
\hline
\end{tabular}
\tablefoot{Listed frequencies were corrected for the stellar line-of-sight Doppler velocity shift, which for this star amounts to approximately $-2.0\:{\rm \mu Hz}$ at $\nu_{\rm max}$.}
\end{table}

\subsection{Frequency modeling}\label{sect:model}
We estimated the mass, $M$, radius, $R$, and age, $t_{\rm age}$, of GJ~570~A and GJ~783~A through frequency modeling (see Table~\ref{tab:param}). A detailed description of the modeling framework will be provided in Costa et al.~(in prep.), and a summary is given below.

We employed a grid of stellar models computed with version r24.08.1 of Modules for Experiments in Stellar Astrophysics \citep[\texttt{MESA};][]{MESA}. Stellar evolutionary tracks were initiated on the pre-main sequence when the central temperature reached $4\times10^{5}\:{\rm K}$ and were evolved until the end of the main sequence, defined as the point where the central hydrogen mass fraction dropped below $10^{-4}$. The grid was constructed by varying the stellar mass, the initial metal fraction, $Z_{\rm init}$, and the mixing-length parameter, $\alpha_{\rm MLT}$, which were treated as free parameters. The mass ranged from 0.60 to $1.15\,M_\odot$ in steps of $0.01\,M_\odot$, $Z_{\rm init}$ ranged from 0 to 0.033 in steps of 0.001 (increasing to 0.002 in parts of the grid), and $\alpha_{\rm MLT}$ ranged from 1.0 to 3.0 in steps of 0.2.

We adopted the solar chemical mixture of \citet{Asplund09}. The initial helium fraction, $Y_{\rm init}$, was determined from a linear helium enrichment law, using the solar-calibrated helium-to-heavy element enrichment ratio from Grid A of \citet{Moedas22}, whose input physics are similar to ours. Convection was treated using a time-dependent formulation based on the model of \citet{Kuhfuss86}. Nuclear reaction rates were taken from the standard \texttt{MESA} compilation, which combines rates from NACRE \citep{NACRE} and JINA REACLIB \citep{JINA_REACLIB}. Opacities were taken from the OPAL tables \citep{OPAL}, supplemented at low temperatures by the tables of \citet{Ferguson05}. Furthermore, we employed the default blended equation of state in \texttt{MESA} and adopted a gray Eddington $T$--$\tau$ relation \citep{Eddington} to treat the atmospheric boundary condition. The models also included element diffusion through gravitational settling \citep{Thoul94}; however, the effects of rotation, convective overshoot, and magnetic fields were neglected. For each model in the grid, we computed oscillation frequencies for modes of angular degree $\ell\!=\!0$--$3$ using the \texttt{GYRE} \citep[v7.2.1;][]{GYRE} stellar oscillation code.

Model optimization was performed with Asteroseismic Inference on a Massive Scale \citep[\texttt{AIMS};][]{AIMS}, which employs Bayesian statistics and a Markov Chain Monte Carlo (MCMC) sampler to identify the best-fitting models within a predefined grid. As observational constraints, we used the mode frequencies listed in Tables~\ref{tab:freq_GJ570A} and \ref{tab:freq_GJ783A}, together with the corresponding effective temperature and metallicity from Table~\ref{tab:param}. Each mode frequency was assigned the same weight as a classical constraint. To correct mode frequencies for the surface effect, we applied the two-term prescription of \citet{BallGizon14}. Although this prescription is widely used for solar-like oscillators, its quantitative validity has not been established specifically for cool K dwarfs at the high $\nu_{\rm max}$ values considered here, and the adopted surface treatment therefore represents a potential source of systematic uncertainty \citep[e.g., see][]{Betrisey23}. For GJ~570~A, the unconstrained \citet{BallGizon14} correction was found to have the opposite sign to that physically expected for the usual surface effect. We therefore restricted the correction to the physically motivated direction, which drove its amplitude close to zero. Given the small number of identified modes for this star and their relatively low radial orders, the available frequencies provide limited leverage on the surface correction; the negligible fitted term should therefore not be interpreted as evidence that the physical surface effect itself is absent. Finally, we accounted for the $\alpha$-enhancement of GJ~783~A by scaling its metallicity following the prescription of \citet{Salaris93}, and used the resulting value as input.

To assess potential systematic effects associated with the optimization procedure, we repeated the analysis using alternative configurations. Specifically, we performed additional runs adopting (i) an alternative weighting scheme, in which the set of mode frequencies and the set of classical constraints were each assigned equal total weight, and (ii) the surface correction of \citet{Sonoi15}. Table~\ref{tab:param} reports separately the statistical uncertainties from the reference run and systematic uncertainties estimated from the dispersion of the inferred parameters across the different optimization configurations.

The radius of GJ~570~A derived from frequency modeling is $4.2\%$ smaller than the interferometric radius (see Table~\ref{tab:param}), corresponding to a $2.6\sigma$ difference. This offset is consistent with the well-known tendency of stellar models to underestimate the radii of K and M dwarfs by typically ${\sim} 5\%$ relative to empirical measurements from eclipsing binaries and interferometry \citep[e.g.,][and references therein]{Boyajian12,FeidenChaboyer12}. Similar discrepancies have also been reported for individual K dwarfs at comparable or higher significance \citep[e.g.,][]{Li25}. To investigate this issue further, Costa et al.~(in prep.) present the first attempt at systematically modeling an ensemble of seismic K dwarfs with published interferometric radii, testing the impact of different physical assumptions.

As GJ~570~A is part of a well-studied multiple system hosting the benchmark T dwarf GJ~570~D, its age has been constrained through several independent diagnostics in the literature. The seismic age derived here, $8.9\pm2.2\,{\rm (stat)}\pm1.1\,{\rm (sys)}\:{\rm Gyr}$, is significantly older than most previous estimates. Early activity and kinematic indicators suggested that the star is older than ${\sim} 2\:{\rm Gyr}$ and belongs to the Galactic disk, implying an approximate upper limit of ${\sim} 10\:{\rm Gyr}$ \citep{Burgasser00}. Subsequent estimates based on chromospheric activity and X-ray emission generally favored younger ages, typically in the range $1$--$5\:{\rm Gyr}$ \citep{Geballe01,Liu07}, while the star's 40-day rotation period implies ${\sim} 6\:{\rm Gyr}$ \citep{Liu07}. Analyses of the brown-dwarf companion GJ~570~D using evolutionary models yielded ages of $3$--$5\:{\rm Gyr}$, with some preference toward the older end of this interval \citep{Saumon06,Phillips20}. The seismic estimate therefore lies above the commonly adopted upper bound of ${\sim} 6\:{\rm Gyr}$. This discrepancy reflects the systematic uncertainties affecting both stellar-model ages and empirical age diagnostics for late-type field stars.

\section{Oscillation amplitudes: observations versus simulations}\label{sect:amp}

\subsection{Empirical mode-amplitude scaling relation}\label{sect:scaling}
The amplitudes of individual oscillation modes are affected by the stochastic nature of the excitation and damping. To measure oscillation amplitudes in a way that is independent of these effects, we adopted the procedure introduced by \citet{Kjeldsen05,Kjeldsen08}. Briefly, the power spectra were heavily smoothed by convolution with a Gaussian kernel of full width at half maximum (FWHM) equal to $4\Delta\nu$, then converted to power density---defined as the power spectrum multiplied by the effective observing duration ($T_{\rm obs}$). We next subtracted the background (white) noise, and finally multiplied the result by $\Delta\nu/c$ and took the square root. Following \citet{Kjeldsen08}, we used $c\!=\!4.09$, which represents the effective number of modes per order for full-disk velocity observations, normalized to the amplitudes of radial modes. This procedure yields a smooth oscillation amplitude envelope, which we refer to as the amplitude per radial mode. We define the maximum amplitude per radial mode, $\varv_{\rm osc,max}$, as the maximum value of this envelope. We obtain $\varv_{\rm osc,max}\!=\!2.57\pm0.52\:{\rm cm\,s^{-1}}$ for GJ~570~A (comparable to the peak amplitude of $2.6\:{\rm cm\,s^{-1}}$ measured for $\epsilon$~Indi) and $4.90\pm0.81\:{\rm cm\,s^{-1}}$ for GJ~783~A. The uncertainties in the oscillation amplitudes were estimated using a bootstrap approach. In each realization, the original power spectrum was multiplied by a synthetic noise spectrum following a $\chi^2$ distribution with two degrees of freedom \citep[][]{Woodard84}, and the oscillation amplitude was remeasured. The dispersion of the resulting amplitude estimates across all realizations was then adopted as the uncertainty.

\begin{figure}[!t]
\centering
\includegraphics[width=0.5\textwidth,trim=0cm 0cm 0cm 0cm,clip]{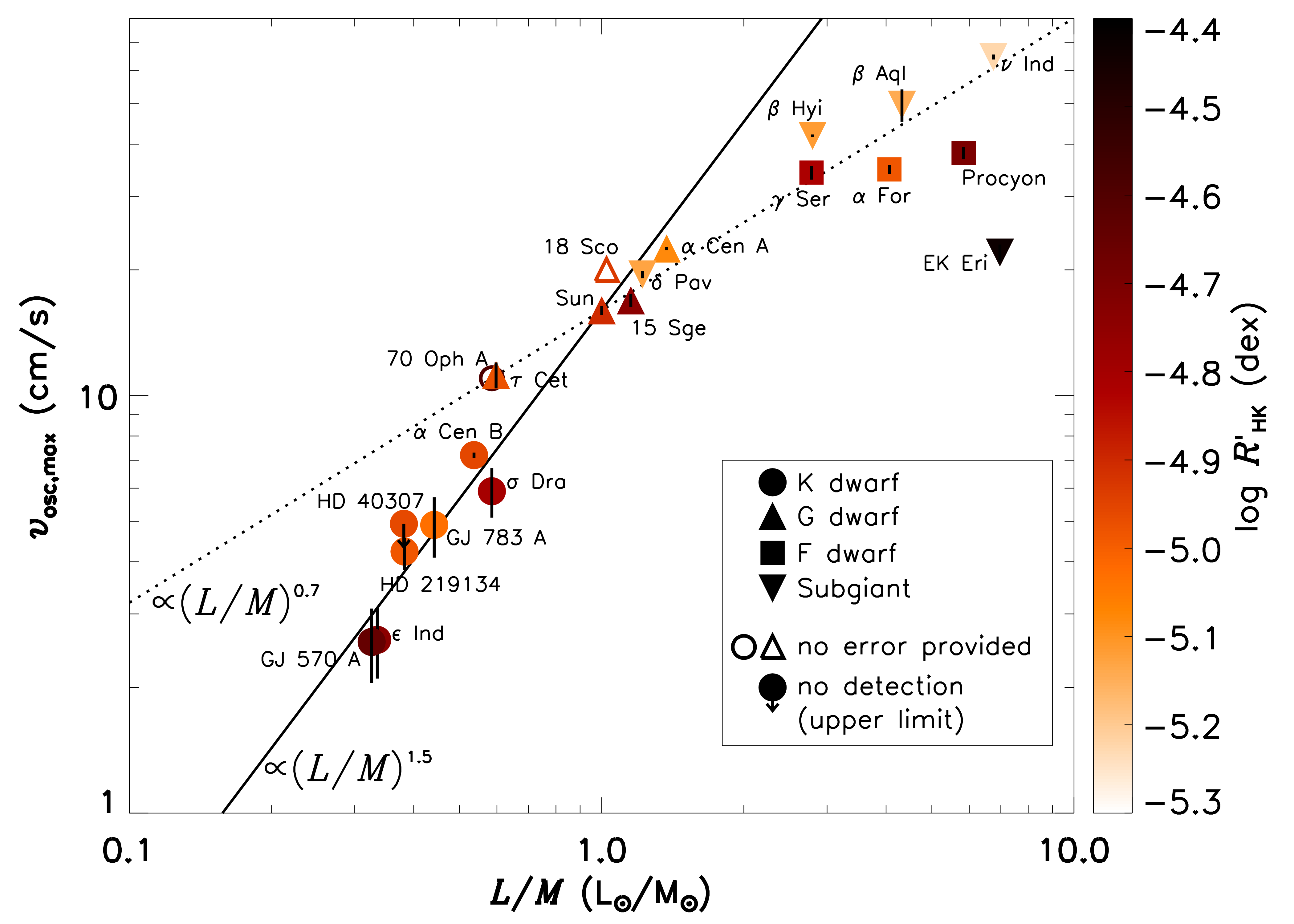}
\caption{Maximum amplitude per radial mode, $\varv_{\rm osc,max}$, as a function of $L/M$ for GJ~570~A, GJ~783~A, the Sun, and a number of FGK dwarfs and subgiants with published Doppler velocity measurements. Data points are color-coded according to the chromospheric emission ratio, $\log R^{\prime}_{\rm HK}$. Two scalings of the mode amplitude are shown, differing in terms of the exponent $s$ ($s\!=\!0.7$, dotted curve; $s\!=\!1.5$, solid curve).}
\label{fig:amp_scaling}
\end{figure}

Figure~\ref{fig:amp_scaling} shows $\varv_{\rm osc,max}$ as a function of $L/M$ (expressed in units of $L_\odot/M_\odot$), color-coded according to the chromospheric emission ratio, $\log R^{\prime}_{\rm HK}$, for a total of 21 stars---GJ~570~A, GJ~783~A, the Sun, and a number of FGK\footnote{Procyon and $\alpha$~Fornacis are moderately evolved F-type stars transitioning from the main sequence toward the subgiant branch, and are therefore often assigned an intermediate luminosity class (IV-V).} dwarfs and subgiants with published Doppler velocity measurements. The amount of smoothing of the power spectrum affects the exact height of the smoothed oscillation amplitude envelope, and hence the estimate of $\varv_{\rm osc,max}$. Measurements for Procyon \citep{Arentoft08}, $\tau$~Ceti \citep{Teixeira09}, EK~Eridani \citep{Bonanno19}, $\epsilon$~Indi \citep{Campante24}, $\sigma$~Draconis \citep{Hon24}, $\beta$~Aquilae \citep{Kjeldsen25}, HD~219134 \citep{Li25}, 15~Sagittae \citep{15Sge}, and the remaining targets \citep[see table~2 of][]{Kjeldsen08} were obtained following the same procedure as described above. There are a few exceptions to this. The value plotted for 70~Ophiuchi~A \citep{CarrierEggenberger06} corresponds to the lower bound on the amplitudes of the highest detected mode peaks and has no associated uncertainty; an estimate of the mode amplitudes for the solar twin 18~Scorpii is given in \citet{Bazot11}, for which no uncertainty is reported; finally, a reanalysis of the ESPRESSO Guaranteed Time Observations (GTO) data for HD~40307 showed no p-mode detection \citep{Campante24}, and so the plotted value is an upper limit. For the Sun, we adopt the determination from \citet{Kjeldsen25} obtained using the same procedure. The underlying velocity amplitude was measured by \citet{Andersen19} from 57 consecutive days of observations with the Stellar Observations Network Group \citep[SONG;][]{SONG26} in mid-2018. \citet{Kjeldsen25} subsequently adjusted this single-epoch measurement to the mean solar activity level using contemporaneous photometric measurements from the Sun photometers (SPM) of the Variability of solar IRradiance and Gravity Oscillations \citep[VIRGO;][]{VIRGO} experiment, combining the green and red SPM channels as a weighted mean. The VIRGO oscillation amplitude during the SONG observations was $3.3\%$ higher than its mean over two solar cycles, leading to the adopted value $\varv^\odot_{\rm osc,max}\!=\!16.0\pm0.4\:{\rm cm\,s^{-1}}$. It is worth noting that the plotted $\varv_{\rm osc,max}$ values are effectively single-epoch determinations and therefore the quoted uncertainties do not account for possible variations induced by magnetic activity cycles \citep[e.g.,][]{Garcia10}; the Sun is a partial exception, in that the value has been corrected to the mean activity level, although it is likewise based on a single-epoch measurement.

\begin{table*}[!t]
\small
\centering
\caption{Basic properties of 3D surface convection simulations. Simulations \texttt{t45g46m00}, \texttt{t46g46m00}, and \texttt{t48g45m05} were designed to model $\epsilon$~Indi, GJ~570~A, and GJ~783~A, respectively. As $T_{\rm eff}$ is an emergent quantity of the 3D simulations that fluctuates over time, both its time-averaged mean and standard deviation are provided.}
\label{tab:3Dsim}
{
\begin{tabular*}{0.8\textwidth}{@{\extracolsep{\fill}}lccc}
\toprule[2pt]
\noalign{\smallskip}
Simulation & $\epsilon$~Indi (\texttt{t45g46m00}) & GJ~570~A (\texttt{t46g46m00}) & GJ~783~A (\texttt{t48g45m05}) \\
\noalign{\smallskip}
\midrule[1pt]
\multicolumn{4}{c}{Standard simulations} \\
\midrule[1pt]
\noalign{\smallskip}
$T_{\rm eff}$ (K) & $4571\pm7$ & $4669\pm8$ & $4854\pm7$ \\
$\log g$ (dex) & 4.62 & 4.57 & 4.55 \\
$[{\rm Fe}/{\rm H}]$ (dex) & 0 & 0 & $-0.5$ \\
Numerical grids & $240^2 \times 260$ & $240^2 \times 260$ & $240^2 \times 254$ \\
Time duration (h) & 10 & 11.67 & 12.5 \\
Sampling interval (s) & 12 & 14 & 15 \\
Horizontal size (Mm) & 3.13 & 3.56 & 4.12 \\
Vertical size (Mm) & 1.79 & 2.06 & 2.47 \\
Vertical grid spacing (km) & 3.8--12.4 & 4.4--14.3 & 4.5--19.4 \\
\noalign{\smallskip}
\midrule[1pt]
\multicolumn{4}{c}{Artificial driving simulations} \\
\midrule[1pt]
\noalign{\smallskip}
Numerical grids & $120^2 \times 158$ & $120^2 \times 158$ & $120^2 \times 158$ \\
Time duration (h) & 50 & 46.67 & 50 \\
Sampling interval (s) & 12 & 14 & 15 \\
Horizontal size (Mm) & 3.13 & 3.56 & 4.12 \\
Vertical size (Mm) & 2.47 & 2.86 & 3.67 \\
Vertical grid spacing (km) & 7.7--24.7 & 8.7--28.5 & 8.9--38.8 \\
\noalign{\smallskip}
\bottomrule[2pt]
\end{tabular*}
}
\end{table*}

The ratio $L/M$ can be expressed as a ratio of atmospheric parameters via $L/M\!\propto\!T_{\rm eff}^4/g$, which we effectively used to compute the quantities along the $x$-coordinate (having adopted $T_{{\rm eff},\odot}\!=\!5777\:{\rm K}$ and $g_\odot\!=\!27402\:{\rm cm\,s^{-2}}$). Estimates of $T_{\rm eff}$ and $\log g$ for GJ~570~A, GJ~783~A, and $\epsilon$~Indi were taken from the spectroscopic analysis in Sect.~\ref{sect:atmo}. For this empirical amplitude-scaling comparison, we retain the spectroscopic surface gravities rather than using the $\nu_{\rm max}$ scaling relation as a seismic constraint to refine $\log g$. This avoids making the inferred $L/M$ values dependent on a solar-calibrated relation whose applicability to cool dwarfs has recently been questioned \citep{Howe20,Hon24,Li25}. For the remaining stars, we adopted spectroscopic $T_{\rm eff}$ and $\log g$ values from the mean PASTEL catalog \citep{PASTEL}; except for Procyon and EK~Eridani, which are not included in that compilation and for which we adopted parameters from the {\it Gaia}--ESO catalog \citep{Gaia-ESO} and \citet{Dall10}, respectively.

In Fig.~\ref{fig:amp_scaling}, we show two scalings of the mode amplitude of the form
\begin{equation}
\begin{aligned}
\varv_{\rm osc,max} \propto (L/M)^s 
&= \varv^\odot_{\rm osc,max} \, \left(\frac{L/L_\odot}{M/M_\odot}\right)^s \\
&= \varv^\odot_{\rm osc,max} \, \left(\frac{T_{\rm eff}^4/T_{{\rm eff},\odot}^4}{g/g_\odot}\right)^s \, ,
\end{aligned}
\label{eq:amp_scaling}
\end{equation}
corresponding to the extrema of the theoretical range in $s$, namely, $s\!=\!0.7$ and $1.5$. A transition can be clearly seen, with decreasing $L/M$, from an $\sim\!(L/M)^{0.7}$ scaling to a steeper $\sim\!(L/M)^{1.5}$ scaling, occurring near the G/K-dwarf boundary at $T_{\rm eff}\!\sim\!5300\:{\rm K}$. This provides compelling support for the trend previously suggested by \citet{Campante24,Hon24,Li25}. In addition, the well-documented overestimation of predicted amplitudes for F-type stars is also apparent in Fig.~\ref{fig:amp_scaling}, with Procyon and $\alpha$~Fornacis lying significantly below the predicted scaling relation \citep[e.g., see][and references therein]{KjeldsenBedding11}.

Measurements of $\log R^{\prime}_{\rm HK}$ for GJ~570~A, GJ~783~A, and $\epsilon$~Indi were taken from Sect.~\ref{sect:chromo} and \citet[for $\epsilon$~Indi]{Campante24}. These are the only stars in the sample for which single-epoch chromospheric activity measurements were used; these measurements are strictly contemporaneous with the asteroseismic data. For the Sun, we adopted a mean chromospheric emission ratio averaged over four consecutive solar cycles \citep{MamajekHillenbrand08}. For the remaining stars, values of $\log R^{\prime}_{\rm HK}$ were drawn from the homogeneous catalog of \citet{GomesdaSilva21} whenever available, or from \citet{BoroSaikia18} otherwise.

K dwarfs tend to exhibit higher levels of chromospheric activity than G dwarfs \citep[e.g.,][]{Jenkins11,GomesdaSilva21}, and increased activity tends to suppress the amplitudes of solar-like oscillations \citep[e.g.,][]{Chaplin00,Chaplin11,Garcia10,Bonanno14,Campante14,Mathur19,Crawford25,Sayeed25}. We examined the relationship between $L/M$ and $\log R^{\prime}_{\rm HK}$ for the stars in our sample, and found a negative correlation that becomes more pronounced when excluding the overactive subgiant EK~Eridani. For this reduced sample, Spearman's rank correlation coefficient is $\rho\!=\!-0.43$ with a two-sided $p$-value of $0.06$, indicating a moderate monotonic trend. A $68\%$ bootstrap confidence interval for the correlation, $[-0.64,-0.17]$, remains entirely negative, indicating that the relationship is not driven by individual stars. This raises the possibility that higher activity levels among K dwarfs may contribute to the transition between amplitude scalings near the G/K-dwarf boundary shown in Fig.~\ref{fig:amp_scaling}.

To assess whether chromospheric activity introduces an additional dependence in the amplitude scaling, we fitted the data with an extended relation of the form \citep[e.g., see][]{Campante14,Crawford25,Sayeed25}
\begin{equation}
\varv_{\rm osc,max}
=
\varv^\odot_{\rm osc,max} \,
\left(\frac{T_{\rm eff}^4/T_{{\rm eff},\odot}^4}{g/g_\odot}\right)^s \,
e^{\,u\,\left(\log R'_{\rm HK}-\log R'_{\rm HK,\odot}\right)} \, ,
\label{eq:amp_scaling_act}
\end{equation}
and compared it to the baseline scaling in Eq.~(\ref{eq:amp_scaling}). The fits were performed in logarithmic space using Monte Carlo propagation of the observational uncertainties in $\varv_{\rm osc,max}$, $T_{\rm eff}$, $\log g$, and $\log R^{\prime}_{\rm HK}$. To account for systematic differences among heterogeneous literature sources, we added in quadrature conservative uncertainty terms of $100\:{\rm K}$ in $T_{\rm eff}$, $0.1\:{\rm dex}$ in $\log g$, and $0.05\:{\rm dex}$ in $\log R^{\prime}_{\rm HK}$, motivated by inter-method and inter-catalog comparisons in the literature \citep[e.g.,][]{Torres12,GomesdaSilva21,PASTEL}. For the two stars lacking reported uncertainties in $\varv_{\rm osc,max}$, we adopted uncertainties of $10\%$. Restricting the analysis to the 19 stars with measured amplitudes (the Sun, which sets the normalization, was excluded) yields a statistically improved fit according to the corrected Akaike Information Criterion (AICc) when including the activity term, together with a reduction in the weighted scatter of the relation. The fitted activity exponent is negative ($u\!=\!-0.88^{+0.39}_{-0.43}$), significant at the $2\sigma$ level and consistent with the expected suppression of oscillation amplitudes with increasing magnetic activity. However, the inclusion of the activity term does not reduce the residual offset between K and G dwarfs, which characterizes the empirical break in amplitude scaling near the G/K-dwarf boundary; the distribution of the corresponding offset difference is symmetric about zero, indicating no meaningful change. We therefore find no clear evidence that magnetic activity contributes to the observed break in the scaling relation near the G/K-dwarf boundary, although a contribution cannot be ruled out given the limited sample size.

\subsection{Tension with 3D surface convection simulations}\label{sect:sim}
From a theoretical perspective, the velocity amplitude of p-mode oscillations can be calculated from non-local, time-dependent theories of convection that consider the interaction between turbulent velocity and pulsations \citep{HoudekDupret15}, or evaluated from 3D compressible hydrodynamical simulations where convection and acoustic oscillations naturally emerge. Various forms of non-local, time-dependent convection models have been used to calculate the linewidths and amplitudes of p modes, with encouraging agreement with observations for the Sun, main-sequence stars, and giants \citep{Balmforth92i,Houdek99,Dupret09}. However, the tunable parameters in these models need to be calibrated against numerical simulations or observational data, which limits their ability to provide independent predictions. In contrast, without the inclusion of free parameters, {\it ab initio} simulations of near-surface convection have been highly successful in quantitatively reproducing the granulation size, spectral line shapes, and center-to-limb variation of solar-type stars \citep{SteinNordlund98,Beeck12,Pereira13}.

\begin{figure*}[!t]
\centering
\includegraphics[height=0.32\textwidth,keepaspectratio]{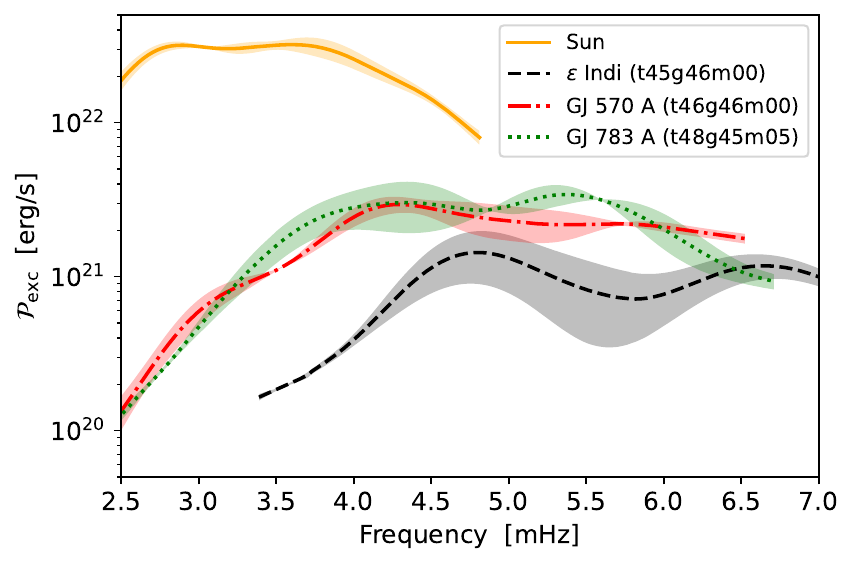}
\includegraphics[height=0.32\textwidth,keepaspectratio]{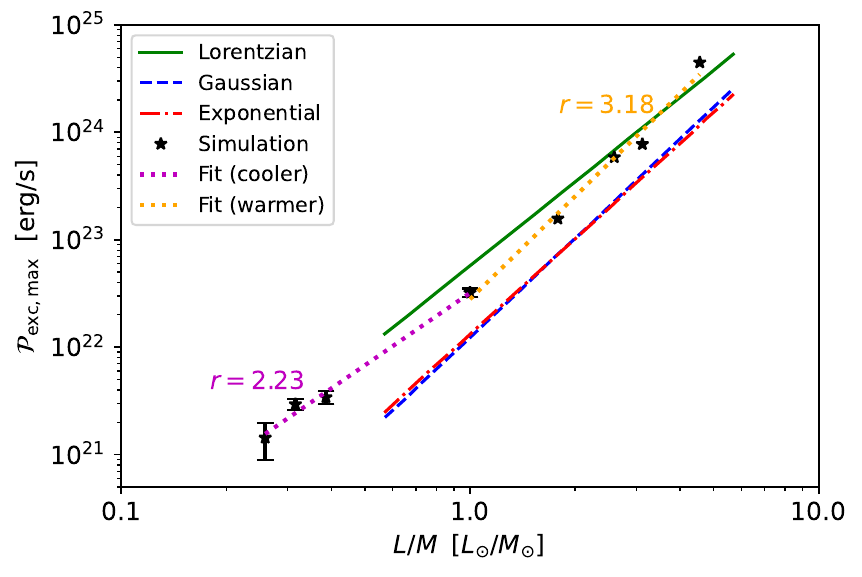}
\caption{\emph{Left:} Energy injection rates, $\mathcal{P}_{\rm exc}$, of radial p modes as a function of frequency for the three K-dwarf models and a solar model \citep{Zhou24}. The curves were smoothed to mitigate the impact of random fluctuations inherent to the simulations. The shaded areas represent the standard deviation, obtained by evaluating $\mathcal{P}_{\rm exc}$ using three different time segments with the same duration. \emph{Right:} Maximum energy injection rate, $\mathcal{P}_{\rm exc,max}$, as a function of $L/M$. Solid, dashed, and dash-dotted lines show the relations from \citet[their fig.~6]{Samadi07}, with the labels ``Lorentzian'', ``Gaussian'', and ``Exponential'' indicating the different analytical turbulence models used in their calculations. Black star symbols represent results from 3D simulations: the three K-dwarf models listed in Table~\ref{tab:3Dsim}, and the solar model together with the warmer models taken from \citet{Zhou20}. Error bars indicate the uncertainties in $\mathcal{P}_{\rm exc,max}$ for the solar and K-dwarf models. Two separate fits to the simulation data are shown: the magenta dotted line is the linear regression of the three K-dwarf models and the solar model, while the orange dotted line corresponds to the linear regression for the warmer models, including the solar model.}
\label{fig:Pexc}
\end{figure*}

To investigate the physical origin of the empirical mode-amplitude scaling for K dwarfs presented in the previous section, we conducted 3D surface convection simulations with the radiative-magnetohydrodynamics code \texttt{Stagger} \citep{NordlundGalsgaard95,Collet18,Stein24} and assessed whether it can be reproduced from first-principles convection modeling. \texttt{Stagger} 3D convection simulations have been extensively validated in the past \citep{Pereira13}. The velocity amplitude of radial p-mode oscillations can be estimated from simulations by quantifying both the energy injection and damping rates, following the methodology described in \citet{NordlundStein01,SteinNordlund01,Zhou20}. In short, the energy injection rate is computed from nonadiabatic pressure fluctuations predicted by the 3D simulations, whereas linear damping rates are quantified through so-called artificial driving simulations \citep[see the appendix of][]{Zhou20}, where a sinusoidal perturbation is applied at the bottom boundary of the simulation domain to artificially excite a radial oscillation to a large amplitude. Mode masses and eigenfunctions, which are required for the calculation of both quantities, are obtained from 1D stellar structure models that are patched to the mean 3D stratification following \citet{Zhou19}. Here, the 1D stellar structure models were computed using \texttt{MESA}, adopting the same setup as described in Sect.~\ref{sect:model}, and the best-fitting model was selected by matching the mean $T_{\rm eff}$ and $\log g$ of the 3D simulation.

\begin{figure*}[!t]
\centering
\includegraphics[width=0.33\textwidth]{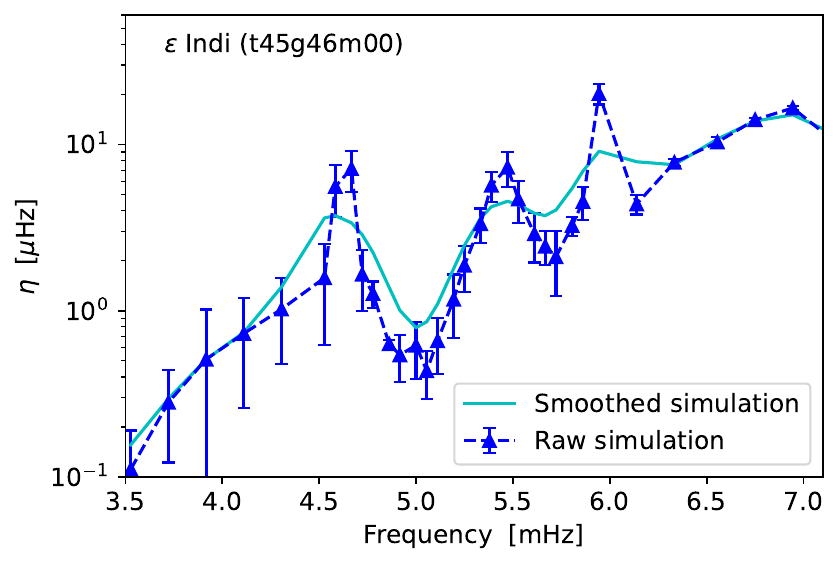}
\includegraphics[width=0.33\textwidth]{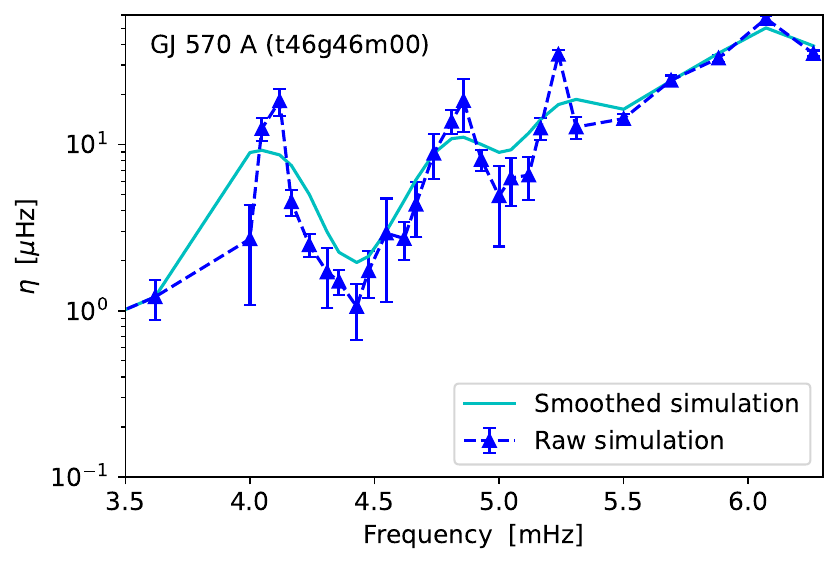}
\includegraphics[width=0.33\textwidth]{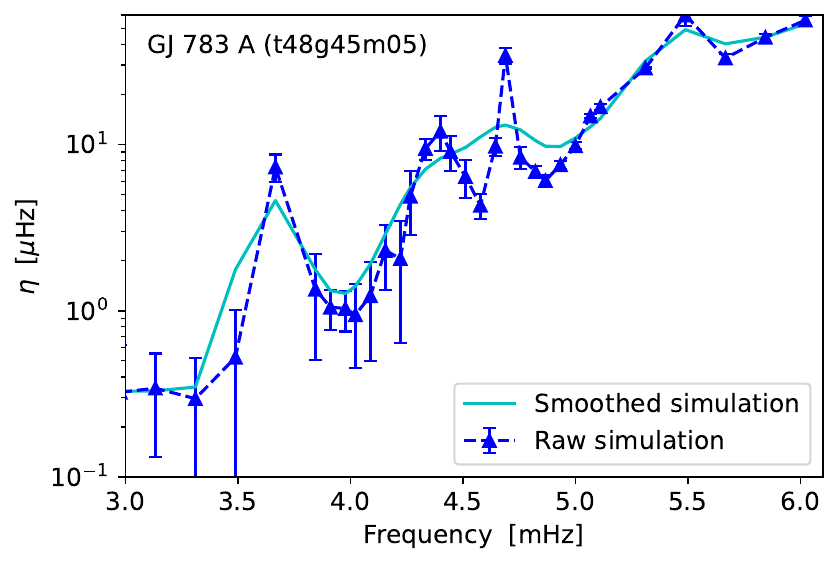}
\includegraphics[width=0.33\textwidth]{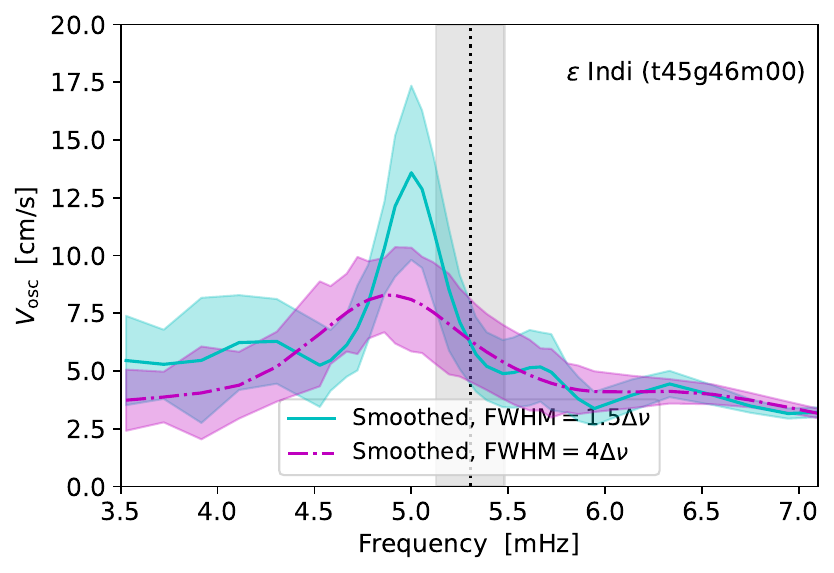}
\includegraphics[width=0.33\textwidth]{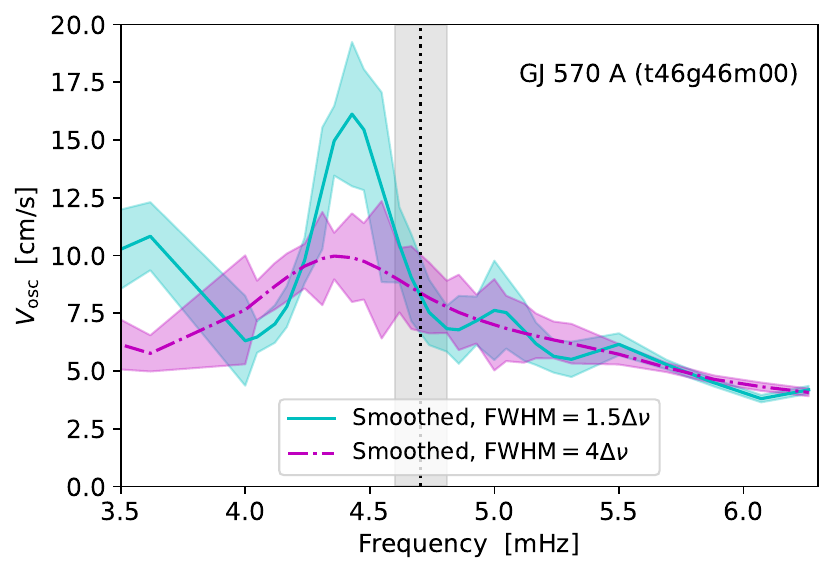}
\includegraphics[width=0.33\textwidth]{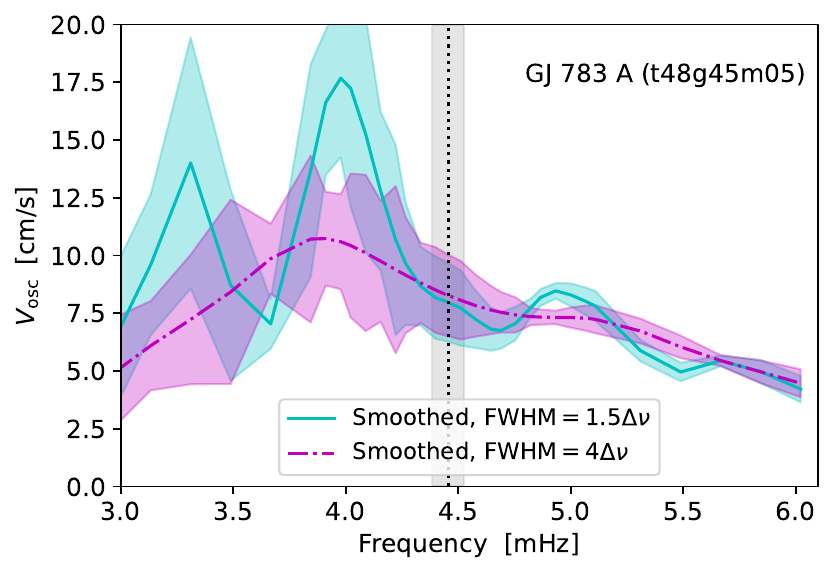}
\caption{\emph{Upper panels:} Linear damping rates, $\eta$, of radial p modes as a function of frequency for the three K-dwarf models (blue triangles connected by dashed lines). Error bars indicate the standard deviation across the nine realizations of the artificial driving simulations at each frequency. The cyan solid curves represent smoothed versions of the raw simulation data. \emph{Lower panels:} Corresponding smoothed (fluid) velocity amplitudes, $V_{\rm osc}$. The cyan solid and magenta dash-dotted curves were obtained by smoothing with Gaussian kernels of FWHM $1.5\Delta\nu$ and $4\Delta\nu$, respectively. The shaded cyan and magenta areas around the respective curves show the corresponding $1\sigma$ uncertainties, propagated from those in the energy injection and damping rates. The $\nu_{\rm max}$ values measured in this work (see Table~\ref{tab:param}) and in \citet[for $\epsilon$~Indi]{Campante24} are indicated by vertical dotted lines, with associated uncertainties shown as gray bands.}
\label{fig:Vosc}
\end{figure*}

Three 3D surface convection simulations were computed, designed to model $\epsilon$~Indi, GJ~570~A, and GJ~783~A (see Table~\ref{tab:3Dsim}). \texttt{Stagger} is equipped with a realistic equation of state and a comprehensive collection of continuum and line opacities. All 3D models were constructed with detailed radiative transfer calculations but without magnetic fields. Basic properties of the simulations used to compute the energy injection rates are summarized in the upper section of Table~\ref{tab:3Dsim}. The mean $T_{\rm eff}$ values of the simulations are consistent with the spectroscopic measurements within uncertainties. The input $\log g$ values were determined from the measured $T_{\rm eff}$ and $\nu_{\rm max}$ according to the $\nu_{\rm max}$ scaling relation \citep{Brown91}. Here, these values are used only to set the atmospheric parameters of the 3D simulations and are not used in the empirical amplitude-scaling relation discussed above. The \citet{Asplund09} metal mixture was adopted in all simulations, consistent with the stellar structure modeling in Sect.~\ref{sect:model}.

Vertically, the simulations span the lower atmosphere, the photosphere, and the superadiabatic convective layers just below the surface, where convective turbulence and nonadiabatic effects responsible for mode excitation and damping are prominent. The highest vertical resolution was applied near optical depth unity to better capture the transition from the optically thick to the optically thin regime. The horizontal size of the simulations was determined based on the typical size of granules for the corresponding stellar parameters \citep{Magic13}. The numerical resolution was halved for the artificial driving simulations (see lower section of Table~\ref{tab:3Dsim}), which were instead extended deeper into the stellar interior, and their temporal coverage was significantly increased, to improve the reliability of the predicted damping rates.

The energy injection rates, $\mathcal{P}_{\rm exc}$, of radial p modes for the three K-dwarf models and a solar model \citep{Zhou24} are shown in the left panel of Fig.~\ref{fig:Pexc}. For each model, the shaded area represents uncertainties estimated from three different time segments with the same duration listed in Table~\ref{tab:3Dsim}. For a given simulation setup, this provides a lower bound on the uncertainty in $\mathcal{P}_{\rm exc}$ arising from the finite duration of the simulation. The energy injection rates exhibit a common frequency dependence: $\mathcal{P}_{\rm exc}$ varies rapidly at low frequencies, reaches a broad plateau encompassing the frequency of maximum oscillation amplitude ($\nu_{\rm max}$), and decreases toward higher frequencies. Regarding the dependence of the energy injection on fundamental stellar parameters, \citet{Samadi07} proposed a power-law relationship between the maximum energy injection rate and the luminosity-mass ratio, $\mathcal{P}_{\rm exc,max}\!\propto\!(L/M)^r$. The corresponding quantities derived from our simulations are shown in the right panel of Fig.~\ref{fig:Pexc}, where $\mathcal{P}_{\rm exc,max}$ is evaluated as the maximum of the smoothed energy injection rate and $L/M$ is computed from the mean $T_{\rm eff}$ and $\log g$ of the 3D model. As shown in the previous section, observed mode amplitudes exhibit a clear transition from an $\sim\!(L/M)^{0.7}$ scaling for warmer stars to a steeper $\sim\!(L/M)^{1.5}$ scaling for cooler stars, occurring near the G/K-dwarf boundary. Motivated by this change in scaling behavior, we performed two separate linear regressions of $\mathcal{P}_{\rm exc,max}$: one for stars with $L/M\!\leq\!L_\odot/M_\odot$, comprising the three K-dwarf models and the solar model, and another for $L/M\!\geq\!L_\odot/M_\odot$, comprising the solar model together with the warmer models studied by \citet{Zhou20}. For the warmer stars, the fitted exponent is $r\!\approx\!3.2$. For the cooler stars, the corresponding exponent, $r\!\approx\!2.2$, is smaller than those predicted by all three turbulence models considered by \citet{Samadi07}. We do not attempt a detailed quantitative interpretation of the fitted exponents, as the estimated $\mathcal{P}_{\rm exc,max}$ values are affected by random fluctuations in the simulations. Nevertheless, the 3D simulations suggest that energy injection into p modes decreases more slowly with decreasing $L/M$ for stars cooler than the Sun, which is at odds with the steeper mode-amplitude decline inferred observationally for these stars.

The stability of an oscillation and its associated energy loss are characterized by the damping rate. For a given oscillation mode, the damping rate is determined by the density fluctuations induced by the oscillation, the pressure fluctuations, and the phase relationship between these two quantities. It is difficult to obtain reliable damping rates from the ``standard'' 3D simulations introduced above, whose density fluctuations contain contributions from both convection and oscillations. To overcome this limitation, we computed linear damping rates for radial oscillations following the method of \citet{Zhou20}, using simulations in which a radial oscillation is continuously driven at the bottom boundary of the simulation domain. This ``artificial driving'' produces a coherent oscillation with large amplitude, for which the density fluctuations predominantly arise from the oscillation. The resulting damping rate should not depend on the magnitude of the perturbation, which requires the amplitudes of the density and pressure fluctuations to scale linearly with the perturbation amplitude. \citet{Zhou20} confirmed that this linear relation holds approximately. To quantify uncertainties associated with the artificial driving method and the finite duration of the simulations, at each frequency we compute damping rates for three perturbation amplitudes combined with three time segments. These artificial driving simulations were repeated at various frequencies to obtain damping rates, $\eta$, as a function of frequency (see upper panels of Fig.~\ref{fig:Vosc}). The error bars represent the standard deviation across the nine realizations, providing a baseline uncertainty for our simulation setup; a more detailed description of the uncertainty evaluation is given in Appendix~\ref{app:solar}.

Once the energy injection and damping rates have been quantified from the simulations, the velocity amplitude can be derived from the condition of energy balance (see lower panels of Fig.~\ref{fig:Vosc}). We note that the velocity amplitude ($V_{\rm osc}$) displayed in Fig.~\ref{fig:Vosc} corresponds to the fluid velocity, whose absolute value is not directly comparable to the observed velocity amplitude ($\varv_{\rm osc}$). The two are instead related by a projection factor (or spatial response function) that captures limb darkening and other geometrical effects associated with disk integration \citep[e.g., see][]{JCDGough82,Kjeldsen08}, which is not explicitly accounted for in the present analysis.

The precise value of $\nu_{\rm max}$ derived from the 3D simulations depends on the adopted smoothing of $V_{\rm osc}$ (see Fig.~\ref{fig:Vosc}). Nevertheless, in line with previous studies \citep{Zhou19,Zhou20}, the simulated $\nu_{\rm max}$ values show broad agreement in scale with observations, lying $\sim\!5$--$10\%$ below the observed values, indicating a systematic offset in excess of the formal observational uncertainties ($\sim\!2$--$3\sigma$ for $\epsilon$~Indi and GJ~570~A, and $>\!3\sigma$ for GJ~783~A). Figure~\ref{fig:Vosc} also reveals the systematic presence of local minima, bracketed by local maxima, in the predicted velocity amplitudes of all three K-dwarf models, most clearly visible in the curves smoothed with the narrower Gaussian kernel \citep[see also][where synthetic p-mode spectra of stars cooler than $\sim\!5400\:{\rm K}$ exhibit two maxima]{Chaplin07}. These amplitude minima correspond to local maxima in the damping rate profiles. The apparent dip above $\nu_{\rm max}$ in the oscillation power of GJ~783~A (see right panel of Fig.~\ref{fig:ps}) is qualitatively consistent with this behavior. However, given the stochastic nature of mode excitation and the limited duration of the observations, we cannot establish whether this feature reflects an intrinsic structure in the oscillation envelope or a particular stochastic realization.

\begin{figure}[!t]
\centering
\includegraphics[width=\columnwidth]{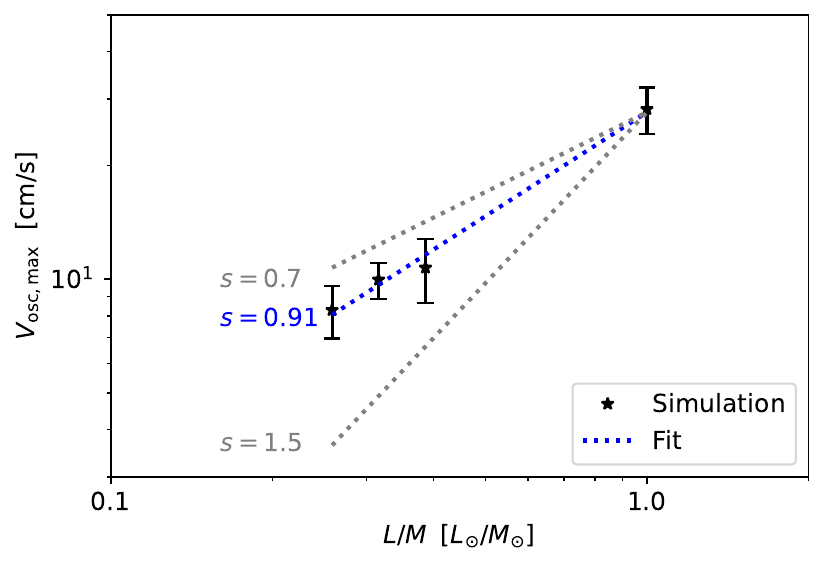}
\caption{Maximum (fluid) velocity amplitudes, $V_{\rm osc,max}$, as a function of $L/M$ for the three K-dwarf models and a solar model \citep{Zhou24}, shown as black star symbols. Error bars indicate uncertainties propagated from those in the energy injection and damping rates. The blue dotted line is the linear regression of $\log V_{\rm osc,max}$ against $\log\, (L/M)$. The gray dotted lines indicate reference scalings with fixed exponents, $s\!=\!0.7$ and $s\!=\!1.5$.}
\label{fig:Vosc_scaling}
\end{figure}

Owing to substantial frequency-dependent fluctuations in both the energy injection and damping rates, only a coarse estimate of the maximum velocity amplitude, $V_{\rm osc,max}$, can be derived. We identified $V_{\rm osc,max}$ as the maximum of the magenta dash-dotted curves in the lower panels of Fig.~\ref{fig:Vosc}, which were obtained by smoothing the raw simulation data with a Gaussian kernel of FWHM $4\Delta\nu$, analogously to the treatment of the observed velocity amplitudes (see Sect.~\ref{sect:scaling}). The resulting estimates of $V_{\rm osc,max}$ for the three K-dwarf models and the solar model of \citet{Zhou24} are shown in Fig.~\ref{fig:Vosc_scaling}. The associated uncertainties were obtained by propagating those in the energy injection and damping rates. A linear regression of $\log V_{\rm osc,max}$ against $\log\, (L/M)$ yields an exponent $s\!\approx\!0.9$ (see Eq.~\ref{eq:amp_scaling}). Although the absolute values of the predicted (fluid) velocities are not directly comparable to the observed values, their dependence on $L/M$ can nevertheless be meaningfully compared with the observationally inferred scaling in Sect.~\ref{sect:scaling}. Strikingly, the 3D convection simulations are unable to recover the steep ($s\!\approx\!1.5$) empirical mode-amplitude scaling of K dwarfs. The mismatch remains robust when the uncertainties in the derived velocities are taken into account. In the most conservative case, adopting the upper limit of the solar velocity and the lower limits for the K-dwarf velocities in the fit yields a slope of $s\!\approx\!1.1$, still well below the empirical value. This result therefore highlights a clear tension between first-principles convection modeling and the observed amplitude behavior of solar-like oscillations in cool dwarfs.

\section{Summary and discussion}\label{sect:discussion}
To place our results in context, we briefly summarize the main findings before discussing the current limitations of our approach and outlining future directions:
\begin{itemize}
\item We carried out high-cadence ESPRESSO campaigns on GJ~570~A (K4~V) and GJ~783~A (K2.5~V), obtaining radial-velocity time series with white-noise levels of $\sim\!1.3$--$1.5\:{\rm cm\,s^{-1}}$, as measured at high frequencies in the amplitude spectra. The corresponding power spectra exhibit an excess consistent with solar-like oscillations in both stars. These detections add to the steadily expanding sample of seismic K dwarfs, strengthening the statistical basis for studies of convection--pulsation coupling.
\item The oscillation power excess is centered slightly below $5\:{\rm mHz}$ in both targets, with individual modes identified for angular degrees $\ell\!=\!0$--$2$ (and $\ell\!=\!3$ in the case of GJ~783~A). In total, 6 modes were measured for GJ~570~A and 24 modes for GJ~783~A. The higher-S/N data for GJ~783~A additionally reveal an apparent dip in the oscillation power above $\nu_{\rm max}$, reminiscent of similar apparent local minima reported for other K dwarfs.
\item From heavily smoothed power spectra, we measured peak radial-mode amplitudes of $2.57\pm0.52\:{\rm cm\,s^{-1}}$ for GJ~570~A and $4.90\pm0.81\:{\rm cm\,s^{-1}}$ for GJ~783~A. These values help reveal a transition from an $\sim\!(L/M)^{0.7}$ scaling to a significantly steeper $\sim\!(L/M)^{1.5}$ dependence, occurring near the G/K-dwarf boundary toward decreasing $L/M$. This provides strong empirical support for the trend previously suggested by \citet{Campante24,Hon24,Li25}. While oscillation amplitudes correlate with chromospheric activity, we find no clear evidence that magnetic activity contributes to the observed break in amplitude scaling, although a contribution cannot be ruled out with the present sample.
\item State-of-the-art 3D surface convection simulations (including explicit calculations of energy injection and linear damping rates) predict a maximum velocity-amplitude scaling $V_{\rm osc,max}\!\propto\!(L/M)^{0.9}$. The simulations thus fail to reproduce the steep empirical $(L/M)^{1.5}$ amplitude scaling observed for K dwarfs, revealing a clear tension between first-principles convection modeling and the measured amplitude behavior in cool dwarfs.
\end{itemize}

The discrepancy between predictions from 3D models and observations suggests that additional physical ingredients and/or limitations of the current modeling assumptions may play an important role, and warrants further investigation. We first consider whether this discrepancy could arise from limitations in the calculation of the energy injection rates. We verified that the vertical extent of the simulation domain, although small compared with the stellar radius, is sufficient for calculating p-mode excitation. For the solar and three K-dwarf models, excluding a small portion near the bottom boundary of the simulation domain has little impact on the resulting energy injection rates, as shown in Fig.~\ref{fig:Pexc-domain}. The 3D simulations suggest that p-mode excitation is concentrated in a thin layer near the stellar surface, especially at high frequencies, consistent with predictions from non-local, time-dependent convection models \citep{Balmforth92iii}.

We next consider the predicted damping rates. We note that linewidths of individual oscillation modes, defined as $\Gamma\!=\!\eta/\pi\!=\!1/(\pi\tau)$, are available for a few main-sequence stars with $T_{\rm eff}\!<\!5500\:{\rm K}$ as part of the {\it Kepler} LEGACY sample \citep[their fig.~20]{Lund17}. For these cool dwarfs, the observed mode linewidths at $\nu_{\rm max}$ are typically around $1\:{\rm \mu Hz}$, with some scatter, translating to damping rates of about $3\:{\rm \mu Hz}$ at $\nu_{\rm max}$. As shown in Fig.~\ref{fig:Vosc}, the simulated $\nu_{\rm max}$ values correspond to the primary local minimum of the damping rate profile for all three 3D models. The smoothed $\eta\,(\nu_{\rm max})$ values are approximately $0.79$, $1.97$, and $1.28\:{\rm \mu Hz}$ for $\epsilon$~Indi (\texttt{t45g46m00}), GJ~570~A (\texttt{t46g46m00}), and GJ~783~A (\texttt{t48g45m05}), respectively. This indicates that our simulations underestimate mode damping by factors of order $1.5$--$4$, and consequently overestimate the oscillation amplitudes. This mismatch could stem from intrinsic uncertainties and limitations of the numerical approach used to estimate $\eta$, or may point to additional damping mechanisms not taken into account in our 3D simulations, such as magnetic fields.

It is reasonable to attribute this tension partly to the absence of magnetic fields in our convection simulations. It is known from both observations \citep{Chaplin00,Chaplin11,Garcia10,Bonanno14,Campante14,Mathur19,Crawford25,Sayeed25} and convection simulations \citep{Jacoutot08,Kitiashvili11} that magnetic fields suppress the amplitudes of p-mode oscillations, a behavior that is also supported by our results (Sect.~\ref{sect:scaling}). Because both $\epsilon$~Indi and GJ~570~A are more active than the Sun, as indicated by their chromospheric emission ratio $\log R^{\prime}_{\rm HK}$, magnetoconvection simulations with appropriate field strengths would plausibly yield lower maximum velocity amplitudes for these stars and, consequently, a steeper fitted slope. Nevertheless, without detailed magnetohydrodynamical modeling, it cannot be determined within the framework of the present study to what extent magnetic suppression accounts for the tension between 3D simulations and observations.

Furthermore, stellar rotation\footnote{Differential rotation also affects the energy injection rates \citep{Biscarrat26}.} may suppress the amplitudes of p modes. Using hydrodynamical simulations of convective stellar interiors, \citet{LeSaux26} found that p-mode amplitudes are anticorrelated with rotation rate, primarily because the energy injection into the modes decreases as the rotation rate increases. In their fastest-rotating model (with $\Omega\!=\!8\,\Omega_{\odot}$, where $\Omega$ is the angular velocity), mode damping is also notably enhanced relative to the non-rotating case. At $r\!=\!0.9\,R$, they predict that the mode power in the solar-rotation-rate model decreases by about $10$--$20\%$ relative to the non-rotating case, corresponding to roughly a $5$--$10\%$ decrease in velocity amplitude. This is smaller than the relative uncertainties in $\varv_{\rm osc,max}$ for the two K dwarfs analyzed in this work. Given that the broader seismic K-dwarf sample considered here is expected to consist of relatively slow rotators, including rotation in the simulations would likely lower the predicted mode amplitudes, but this effect is not expected to be the dominant factor behind the discrepancy in the amplitude scaling.

Another complication arises from the difference between $V_{\rm osc,max}$ derived from the 3D simulations and the observed RV oscillation amplitudes. The former correspond to velocities evaluated at a fixed geometrical layer close to Rosseland optical depth unity, whereas the latter represent velocities sampled over the effective line-formation heights of the spectral lines used in the RV extraction. Although both refer to layers within the photosphere, many of the lines contributing to RV measurements are formed at smaller continuum optical depths, i.e., geometrically higher atmospheric layers, than the $\tau_{\rm Ross}\!\approx\!1$ surface. Furthermore, the formation heights of spectral lines in the Sun differ from those in K dwarfs, implying that the fitted slope would change if synthesized RVs from detailed line-formation calculations were used instead of velocities at a fixed depth. For \ion{Fe}{i} lines, which constitute a major set of lines in the ESPRESSO cross-correlation masks \citep{ESPRESSO}, the same transition, at fixed iron abundance, generally exhibits a larger equivalent width in a K dwarf than in the Sun. This is primarily due to the higher fraction of neutral iron at lower effective temperatures, which increases the line opacity. As a result, these lines form at smaller continuum optical depths, corresponding to geometrically higher layers within the photosphere, in K dwarfs. Because oscillation velocity amplitudes increase with height in the atmosphere \citep[e.g.,][]{Isaak89,Baudin05,Houdek06,Luhn25}, using velocities evaluated near $\tau_{\rm Ross}\!\approx\!1$ likely underestimates the observed RV oscillation amplitudes, with the effect being stronger at lower $L/M$. It follows that incorporating line-formation effects would preferentially increase the modeled amplitudes of K dwarfs, thereby further flattening the inferred scaling of $V_{\rm osc,max}$ with $L/M$ and yielding an exponent even smaller than that obtained from velocities evaluated at a fixed depth. An additional factor affecting the RV oscillation amplitudes is that different instruments rely on different spectral lines for RV measurements. Even for a given instrument, the set of lines used can vary from star to star. Consequently, it is not straightforward, even at a qualitative level, to assess the impact of adopting velocities from a fixed depth on the fitted slope $s$. A detailed synthesis of RV signals from simulations, using a method that resembles the observational procedure, is required to derive velocity amplitudes that are fully comparable with observations. Such an analysis lies beyond the scope of the present work but will be addressed in future studies.

\begin{acknowledgements}
Based on observations collected at the European Southern Observatory (ESO) under programs 0113.D-0255(A) (GJ~783~A; PI: Campante) and 0113.D-0255(B) (GJ~570~A; PI: Campante). The simulations were partly performed on resources provided by Sigma2, the National Infrastructure for High-Performance Computing and Data Storage in Norway. This research was supported by computational resources provided by the Australian Government through the National Computational Infrastructure (NCI) under the National Computational Merit Allocation Scheme and the ANU Merit Allocation Scheme (project y89). This work was supported by Funda\c c\~ao para a Ci\^encia e a Tecnologia (FCT) through research grant \href{https://doi.org/10.54499/UID/04434/2025}{UID/04434/2025}. This work was supported by national funds through FCT under project \href{https://doi.org/10.54499/2024.15303.PEX}{2024.15303.PEX}. TLC is supported by FCT in the form of a work contract (\href{https://doi.org/10.54499/2023.08117.CEECIND/CP2839/CT0004}{2023.08117.CEECIND/CP2839/CT0004}). YZ acknowledges support from the European Union's Horizon 2020 research and innovation program under Marie Skłodowska-Curie grant agreement no.~101150921. YL and DH acknowledge support from NASA (80NSSC25K7904) and its Interdisciplinary Consortia for Astrobiology Research (NNH19ZDA001N-ICAR) under award no.~19-ICAR19 2-0041. VA is supported by FCT in the form of a work contract (\href{https://doi.org/10.54499/2023.06055.CEECIND/CP2839/CT0005}{2023.06055.CEECIND/CP2839/CT0005}). JRL and MSL are supported by research grant 42101 from VILLUM FONDEN. TRB is supported by the Australian Research Council through Laureate Fellowship FL220100117. EC acknowledges support from the research grant “Unveiling the magnetic side of the Stars” (PI: A.~Bonanno) funded under the INAF national call for Fundamental Research 2023. MSC is supported by FCT in the form of a work contract (\href{https://doi.org/10.54499/2023.09303.CEECIND/CP2839/CT0003}{2023.09303.CEECIND/CP2839/CT0003}). This work was funded by the European Union (ERC, FIERCE, 101052347). Views and opinions expressed are, however, those of the authors only and do not necessarily reflect those of the European Union or the European Research Council. Neither the European Union nor the granting authority can be held responsible for them. The authors thank Anish Amarsi and Thomas Nordlander for valuable comments and fruitful discussions.
\end{acknowledgements}

%

\bibliographystyle{aa} 
\bibliography{biblio.bib} 


\listofobjects


\begin{appendix}

\section{Predicted solar velocity amplitudes}
\label{app:solar}

\begin{figure}[!t]
\centering
\includegraphics[width=\columnwidth]{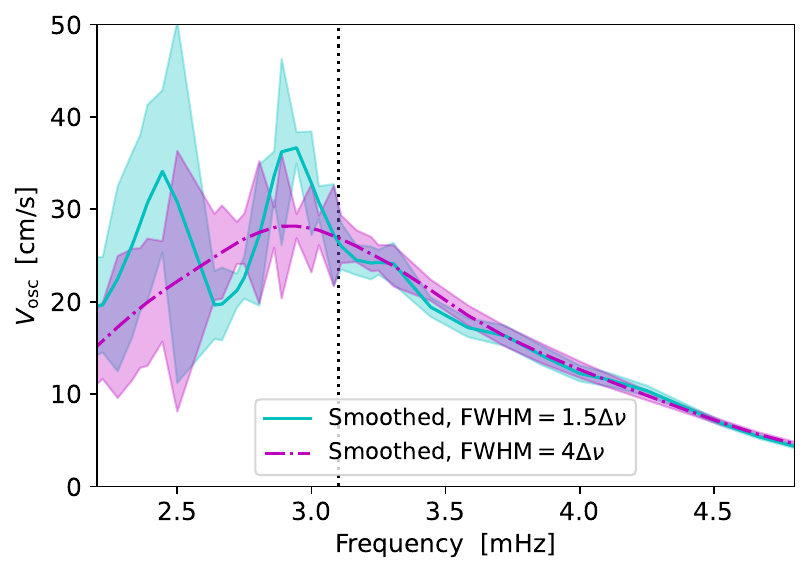}
\caption{Smoothed velocity amplitudes predicted from a \texttt{Stagger} solar simulation. The cyan solid and magenta dash-dotted curves were obtained by smoothing with Gaussian kernels of FWHM $1.5\Delta\nu$ and $4\Delta\nu$, respectively. The shaded cyan and magenta areas around the respective curves show the corresponding $1\sigma$ uncertainties, propagated from those in the energy injection and damping rates. The solar $\nu_{\rm max}$ value measured in \citet{Kjeldsen08} is indicated by a vertical dotted line.}
\label{fig:solar-V}
\end{figure}

The maximum solar velocity amplitude shown in Fig.~\ref{fig:Vosc_scaling} corresponds to the maximum of the magenta dash-dotted theoretical velocity curve in Fig.~\ref{fig:solar-V}. The theoretical velocity amplitudes of radial p modes are estimated assuming energy balance between mode excitation and damping, with the energy injection and damping rates computed as described in Sect.~\ref{sect:sim}, based on the solar simulations summarized in table~1 of \citet{Zhou24}.

The baseline uncertainties in the energy injection rates, damping rates, and resulting velocity amplitudes are estimated from multiple realizations of the same simulation setup. These should be regarded as lower bounds on the corresponding uncertainties. Three realizations of the energy injection rates are computed from three different time segments of the simulation, with identical durations and numerical settings. For the damping rates, which are derived from artificial driving simulations in which a sinusoidal perturbation is applied at the bottom boundary of the simulation domain, the amplitude of this perturbation introduces an additional source of uncertainty. We therefore combine three time segments with three perturbation amplitudes at each frequency, yielding nine realizations of the damping rate. The scatter across these nine realizations provides an estimate of the uncertainty arising from both the finite duration of the simulations and the artificial driving method. Finally, since $V_{\rm osc}\!\propto\!\sqrt{\mathcal{P}_{\rm exc}/\eta}$, the relative uncertainties in the mean energy injection and damping rates are propagated to the velocity amplitudes in quadrature:
\begin{equation}
    \sigma_V = \frac{V}{2}\sqrt{\left(\frac{\sigma_{\mathcal{P}}}{\overline{\mathcal{P}}}\right)^2 + \left(\frac{\sigma_{\eta}}{\overline{\eta}}\right)^2} \, ,
\end{equation}
where $\sigma$ denotes the standard deviation across the realizations, and the bar symbol denotes the corresponding mean. This procedure yields the shaded areas in Fig.~\ref{fig:solar-V}, which represent the $1\sigma$ uncertainties in the derived velocity amplitudes.

\section{Dependence of $\mathcal{P}_{\rm exc}$ on simulation depth}
\label{app:excitation}

The energy injection rates of radial p modes are computed following \citet{Zhou24}:
\begin{equation}
    \label{eq:Pexc}
    \begin{aligned}
    \mathcal{P}_{\rm exc}(\omega) =&
    \frac{\omega^2 A_{\rm box}}{8 \mathcal{T}_{\rm tot} E_0} \left[ 
    \left( \int_{r_{\rm bot}}^{r_{\rm top}} \frac{\partial \xi_r}{\partial r} 
    \mathfrak{Re}\left\lbrace \mathcal{F}[\delta \bar{P}_{\rm nad}] \right\rbrace {\rm d}r \right)^2 \right.
    \\
    +& \left. \left(\int_{r_{\rm bot}}^{r_{\rm top}} \frac{\partial \xi_r}{\partial r} 
    \mathfrak{Im}\left\lbrace \mathcal{F}[\delta \bar{P}_{\rm nad}] \right\rbrace {\rm d}r \right)^2 \right] \, ,
    \end{aligned}
\end{equation}
where $\mathfrak{Re}$ and $\mathfrak{Im}$ denote the real and imaginary parts, respectively. The terms $\omega$, $A_{\rm box}$, and $\mathcal{T}_{\rm tot}$ are the angular frequency, horizontal area, and total duration of the simulation, respectively. $\xi_r$ and $E_0$ are the radial-mode eigenfunction and mode kinetic energy per unit surface area, respectively, while $\mathcal{F} [\delta \bar{P}_{\rm nad}]$ is the Fourier transform of the horizontally averaged non-adiabatic pressure fluctuation induced by convection. We refer the reader to \citet{Zhou24} for a detailed discussion of the physical interpretation and numerical evaluation of $\mathcal{P}_{\rm exc}$.

\begin{figure*}[!t]
\centering
\includegraphics[width=0.48\textwidth]{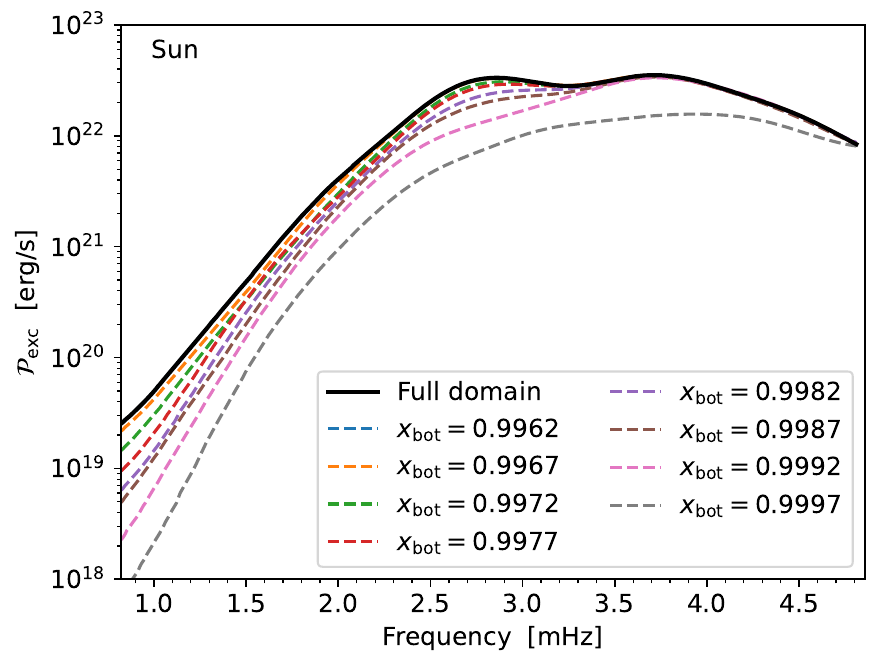}
\includegraphics[width=0.48\textwidth]{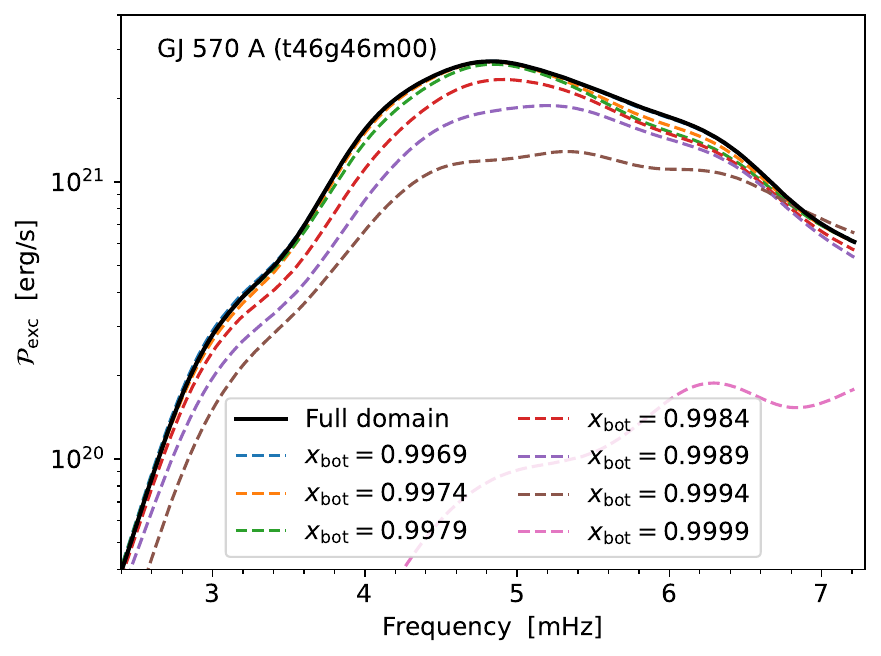}
\caption{\emph{Left panel:} Energy injection rates for the solar model. Dashed curves correspond to different lower integration limits, expressed in terms of fractional radius as $x_{\rm bot}\!=\!r_{\rm bot}/R$, in Eq.~(\ref{eq:Pexc}). The result obtained using the full simulation domain is shown by the black solid curve for reference. \emph{Right panel:} Same as the left panel, but for the model corresponding to GJ~570~A.}
\label{fig:Pexc-domain}
\end{figure*}

For the simulations shown in Fig.~\ref{fig:Pexc}, the integration in Eq.~(\ref{eq:Pexc}) is carried out from the bottom boundary of the simulation domain. To examine whether the vertical extent of the simulations is sufficient to provide a reliable estimate of mode excitation, we evaluate Eq.~(\ref{eq:Pexc}) using different lower integration limits, $r_{\rm bot}$, while keeping the upper limit, $r_{\rm top}$, fixed. Energy injection rates computed using different lower integration limits are shown in Fig.~\ref{fig:Pexc-domain} for the solar model (left panel) and the model corresponding to GJ~570~A (right panel). In both cases, lower integration limits of $r_{\rm bot}\!\lesssim\!0.998\,R$ yield energy injection rates that agree well with those obtained using the full simulation domain, indicating that the contribution from deeper layers is small.

\end{appendix}

\end{document}